\documentclass[onecolumn]{article}
\usepackage[english]{babel}

\usepackage[utf8]{inputenc}
\usepackage[pdftex]{graphicx}
\usepackage{authblk}
\usepackage{dcolumn}
\usepackage{amsmath}
\usepackage{longtable}
\usepackage{tabularx}
\usepackage{setspace}
\usepackage{multirow}
\usepackage{mathtools}
\usepackage{tikz}
\usepackage{lscape}
\usepackage{amssymb}
\usepackage{url}
\usepackage{arydshln}
\usepackage{hyperref}
\usepackage[misc]{ifsym}

\newcommand{\beginsupplement}{%
        \setcounter{table}{0}
        \renewcommand{\thetable}{S\arabic{table}}%
        \setcounter{figure}{0}
        \renewcommand{\thefigure}{S\arabic{figure}}%
        \setcounter{equation}{0}
        \renewcommand{\theequation}{S\arabic{equation}}%
        \setcounter{section}{0}
        \renewcommand{\thesection}{S\arabic{section}}%
     }

\usepackage{csquotes}
\MakeOuterQuote{"}
\usepackage{lscape}

\usepackage[font=small]{caption}

\RequirePackage[twoside,letterpaper,includeheadfoot,layoutsize={8.125in,10.875in},layouthoffset=0.1875in,layoutvoffset=0.0625in,left=38.5pt,right=43pt,top=43pt,bottom=32pt,headheight=0pt,headsep=10pt,footskip=25pt]{geometry}
\usepackage{array}
\newcommand{\PreserveBackslash}[1]{\let\temp=\\#1\let\\=\temp}
\newcolumntype{C}[1]{>{\PreserveBackslash\centering}p{#1}}
\newcolumntype{R}[1]{>{\PreserveBackslash\raggedleft}p{#1}}
\newcolumntype{L}[1]{>{\PreserveBackslash\raggedright}p{#1}}

\title{Citizens at the forefront of the constitutional debate: Participation determinants and emergent content in Chile}

\author[1 \Letter]{\small M. P. Raveau}
\author[4,5]{\small J. P. Couyoumdjian}
\author[6]{\small C. Fuentes-Bravo}
\author[1]{\small C. Rodriguez-Sickert}
\author[1,2,3 \Letter]{\small Cristian Candia}

\affil[1]{Centro de Investigación en Complejidad Social (CICS), Facultad de Gobierno, Universidad del Desarrollo, Santiago, Chile.}

\affil[2]{\small Kellogg School of Management, Northwestern University, Evanston, IL 60208}

\affil[3]{Northwestern Institute on Complex Systems (NICO), Northwestern University, Evanston, IL 60208}

\affil[4]{Centro de Políticas Públicas, Facultad de Gobierno, Universidad del Desarrollo}

\affil[5]{Facultad de Economía y Negocios, Universidad del Desarrollo}

\affil[6]{Instituto de Argumentación, Universidad de Chile, Santiago, Chile.} 

\affil[ \Letter ]{Corresponding Authors: mraveaum@udd.cl, cristian.candia@kellogg.northwestern.edu}

\date{\today}   

\begin{document}
\maketitle

\begin{abstract}
In the past few decades, constitution-making processes have shifted from closed elite writing to incorporating democratic mechanisms. Yet, little is known about democratic participation in deliberative constitution-making processes. Here, we study a deliberative constituent process held by the Chilean government between 2015 and 2016. The Chilean process had the highest level of citizen participation in the world ($204,402$ people, i.e., $1.3\%$ of the population) for such a process and covered $98\%$ of the national territory. In its participatory phase, people gathered in self-convoked groups of 10 to 30 members, and they collectively selected, deliberated, and wrote down an argument on why the new constitution should include those social rights. To understand the citizen participation drivers in this volunteer process, we first identify the determinants at the municipality level. We find the educational level, engagement in politics, support for the (left-wing) government, and Internet access increased participation. In contrast, population density and the share of evangelical Christians decreased participation. Moreover, we do not find evidence of political manipulation on citizen participation. In light of those determinants, we analyze the collective selection of social rights, and the content produced during the deliberative phase. The findings suggest that the knowledge embedded in cities, proxied using education levels and main economic activity, facilitates deliberation about themes, concepts, and ideas. These results can inform the organization of new deliberative processes that involve voluntary citizen participation, from citizen consultations to constitution-making processes.




\end{abstract}



\maketitle

\section*{Introduction}
\label{sec:intro}

"The punishment which the wise suffer who refuse to take part in the government is to live under the government of worse men," Plato wrote in The Republic\footnote{Variation of the original phrase "But the chief penalty is to be governed by someone worse if a man will not himself hold office and rule." Book 1, 346–347, made by Ralph Waldo Emerson, Society and Solitude (1870)}. This sentence wisely illustrates the relevance of political participation in all its extent. From Plato's epoch, democracy has evolved at incorporating new ways of citizen participation. From Solon asking people to swear for respecting his rules to recent deliberative processes, where people debate about relevant topics, and the conclusions are used as inputs in executive decision-making or even to write a new Carta Magna. 

In the last decades, political participation studies have been mainly focused on voter turnout, engagement in political parties, and civil disobedience in the light of participation determinants such as gender \cite{pachon2012, schlozman1999, pnud}, age \cite{wolfinger1980, highton2001, contreras}, education and income\cite{pnud,verba, corvalan}, and social capital \cite{coleman1988, ladua, klesner, bargsted, mccull, campbell, muller1970, hofferth}. Yet, we still lack a full understanding of democratic and deliberative constitutional-making processes. Recently, the long tradition of elite-class-produced constitutions shifted to a democratic constitution-making process because of it lacked legitimacy \cite{usip}. This new paradigm incorporates citizen participation ex-ante, ex-dure, and ex-post the constitutional text finalization \cite{banks2007, ginsburg2009, usip}, which represents an epitome for political participation. Thus, Plato's quote captures an emergent political trend, citizens at the forefront of the constitutional debate. 

\begin{figure}[!t]
\centering
\includegraphics[width=.99\textwidth]{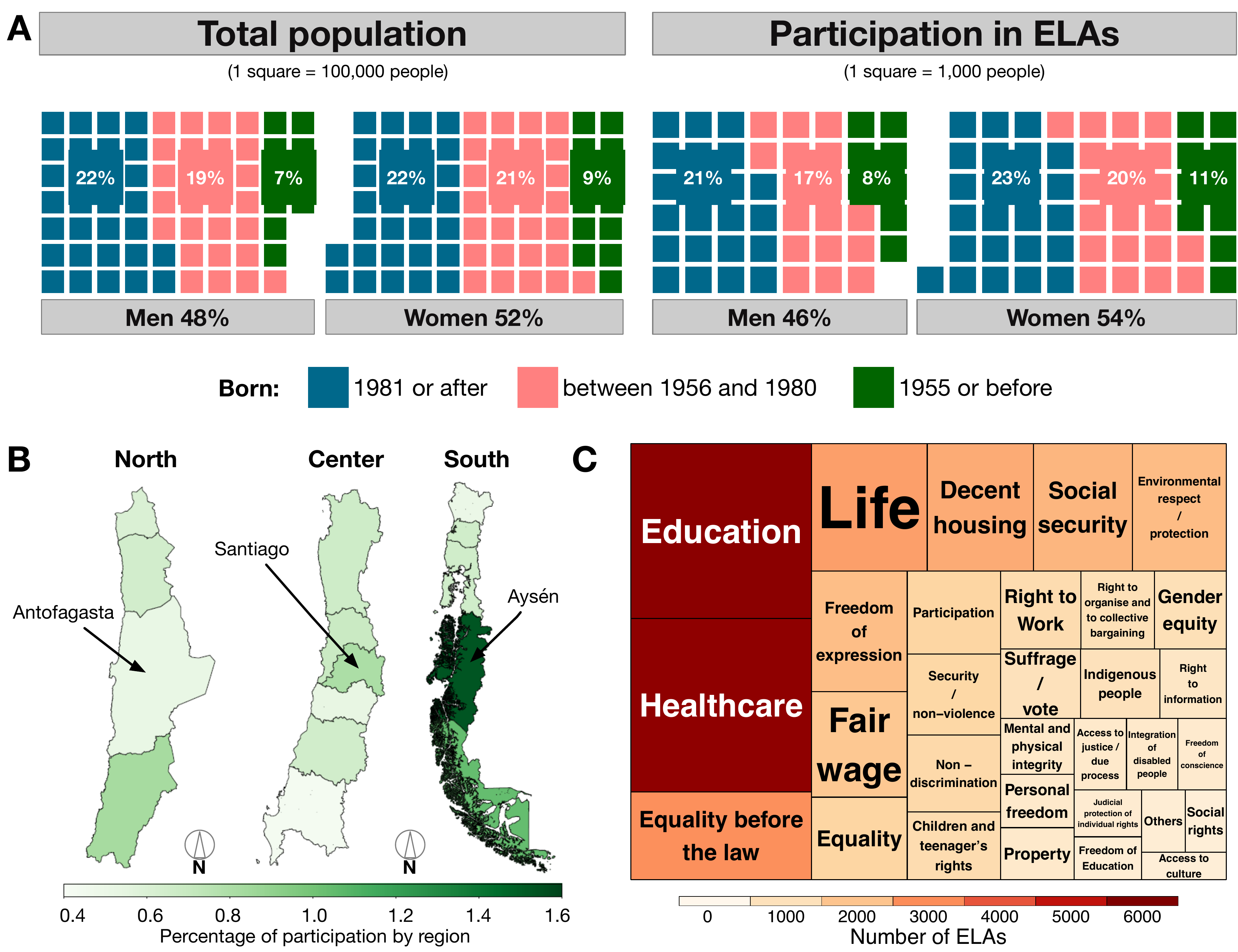}
\caption{A) Composition of national population from Census 2017 data and of citizen participation in self-convoked encounters (ELAs), by gender and generational cohorts (colors). Participation distribution does not differ significantly from the population distribution. The contingency tables and chi-squared test results can be found in the Supplemental Material (Tables~\ref{tab:age_chi} and \ref{tab:sex_chi}). B) Percentage of the population participating in ELAs per region in Chile. From left to right, the northern macro-zone, the center macro-zone, and the southern macro-zone.  C) What constitutional rights do people select at the country level? The tree-map depicts the most selected constitutional rights in all the self-convoked encounters, ELAs. Color and size represent the number of ELAs in which those concepts were selected.}
\label{fig:edad}
\end{figure}

In October 2015 the government of Chile proposed a new constitution-making process, whose final aim is to generate a new social contract. In this process, citizens are at the forefront of the constitutional debate as active participants, generating input for the proposed new constitution. According to an OECD report,\cite{ocde} the case of Chile is unprecedented because of its high rate of participation ($1.13\%$), compared to other experiences, and coverage of $98\%$ of the territory (Fig. \ref{fig:edad}B). Moreover, in terms of gender and age, the participation distribution in the Chilean deliberative process does not differ significantly from the population distribution, suggesting that it represents citizens at the country-level for these two dimensions (Fig. \ref{fig:edad}A). 

The constitutional process started with sessions of civic education for citizens, provided by the government. Then, the participatory phase took place, divided into four stages: i) online individual questionnaire;  ii) local deliberative self-convoked encounters (in Spanish \textit{Encuentros Locales Autoconvocados}, henceforward ``ELAs''); iii) dialogues at the province level; and iv) dialogues at the regional level. At each stage, citizens were asked to debate four constitutional dimensions: i) constitutional principles and values; ii) rights, iii) duties, and iv) institutions. 

Given the increasing demands on social rights in Chile, we specifically focus on the local self-convoked encounters in the dimension of \textit{rights}. $8,113$ ELAs were held during the participatory phase, with a total of $106,412$ participants \cite{memoria}, and each ELA consisted of between 10 to 30 people, all over 14 years old \cite{guiacabildo}. For each of the four constitutional dimensions, participants collectively had to select seven constitutional concepts from a list provided by the government or provide new constitutional concepts after group deliberation (Fig. \ref{fig:edad}C). For each chosen or provided concept, they had to write down a short argument explaining why this concept should be included in the new constitution. The result of this consultation would serve as an input to the elaboration of a new Constitution proposal written by the executive power \cite{ocde}. The proposal was submitted to Congress in March 2018. 

Unlike voting, this form of participation is intrinsically social and requires higher commitment, since the ELA required a quorum of 10 people to be valid. Participation in Chilean's constituent process also represents a different type of democratic exercise, a deliberative democracy. Here, we study the determining factors of citizen participation in ELAs by setting up various regression models at the municipality-level. We included socio-demographic and political variables, as well as social capital indicators from different data sources, such as census, the Electoral Service, and the national municipal information system, among others. Regarding the influence of socio-demographic variables, one important finding is that engagement in politics and support for the government increases participation, which suggests that citizen involvement in the constitutional process may have been ideologically driven. Then, we analyze the effects of the citizen participation determinants and other relevant variables on the selection of constitutional rights. Finally, using structural topic modeling, we identify the latent topics in the argument texts. Then, we explore the changes of content for each topic for cities with different characteristics, showing that the emergent content can be ideologically differentiated. 

\subsection*{Participation in constitution-making processes}

In 1787, the United States of America originated a tradition of elite group constitution-making. However, it is now believed that democratic constitutions should be created and adopted through democratic processes because elite-made constitutions suffer from a lack of legitimacy,  i.e., the sense that the constitution represents the people it governs \cite{elster1993,ginsburg2009,usip}. This idea relies on the belief that the sense of ``ownership'' that comes from sharing authorship, enhances the understanding, respect, and support of the constraints that the constitution imposes. Yet, these assumptions have been subject to only limited study, and the precise relationship between participation and its desirable outcomes remains insufficiently studied \cite{ginsburg2009}.

Recent works on evaluation of constitution-making have investigated its effects on democratic governance. A 12-country study, commissioned by the Institute for Democracy and Electoral Assistance (IDEA), found that ``more representative and inclusive constitution building processes resulted in constitutions favoring free and fair elections, greater political equality, more social justice provisions, human right protections, and stronger accountability mechanisms''\cite{samuels2006}. Later, it was found that public referenda make constitutions more likely to include every category of right, and to provide for universal suffrage, a secret ballot, a referendum process in ordinary government, and a public role in approving constitutional amendments \cite{ginsburg2009}.

Recent constitutions have redefined the long tradition of elite constitution-making, incorporating citizen participation before, during, and after the text is finalized.  This new form of constitution making is supported by numerous international organizations and entities, such as the European Union, the Commonwealth Human Rights Initiative, the United States Institute of Peace, and the Centre for Democracy and Development \cite{banks2007}. Models of participation differ considerably, not only among countries, but within a country. The modal form of participation in constitutional design is the approval by referendum of the final document as a whole \cite{ginsburg2009}. Other forms of participation include (i) elections for constitutions-making assemblies, (ii) civic education and media campaigns using newspapers, radio, television, web sites, or public meetings; (iii) prior agreement on broad principles as a first phase of constitution making, and (iv) an interim constitution to create space for longer term democratic deliberation \cite{ginsburg2009,usip}.

\subsection*{Political participation}

Most research on political participation has been focused on voting turnout,  engagement in political parties and public debates, and civil disobedience, and in recent decades, voter turnout has led research in the field of political science. While participation is, in general, facilitated by socioeconomic resources and the institutional context, voting is particularly influenced by the competitive context in which elections take place, and by institutional factors, such as registration rules, electoral system, and linkage between parties and social groups \cite{powell}. Here, we introduce some of the main findings on political participation, but, while this form of participation will be similar to constituent processes in some aspects, it likely differs in others. Examining features of the citizens helps us understand differences in political participation.

\textit{Gender}: Latin American women are less likely to be involved in political activities, especially in  protest demonstrations and party affiliation. \cite{pachon2012, schlozman1999}, however, this gender gap in political participation decreases as female participation in labor market increases \cite{schlozman1999}. Moreover, a recent study on electoral turnout in Chile shows that for almost all age groups voter turnout rates for women exceeds voter turnout rates for men \cite{pnud} (See Supplementary note \ref{SN1}).

\textit{Age}: The level of participation is relatively low during early adult life, then gradually increases among middle-aged voters, and softly declines with old age \cite{wolfinger1980}. Particularly in Chile, people who reached the voting age after the end of the military dictatorship of general Pinochet, in 1988, show lower enrollment and participation rates \cite{contreras} (See Supplementary note \ref{SN1}).

\textit{Education and income}: Income and education level have a positive and significant effect on political participation \cite{verba, pnud}. In Chile, the young electorate is strongly income-biased, which causes voter registration rates in the upper class to double those of lower class \cite{corvalan} (See Supplementary note \ref{SN1}).

\textit{Social capital}: Social capital refers to the social relationships that facilitate action. The production of social capital encourages citizens to become more engaged in a broader range of traditional political activities \cite{ladua, klesner}. Social capital has been proxied using population density and rural population. Population density has shown a negative impact on social capital and social organization for men \cite{mccull}, which would negatively affect voter turnout. The urban and rural residency has mixed conclusions \cite{campbell,muller1970, hofferth}. Particularly for Chile, population density has shown a negative effect, and the rural population has a non-significant impact on electoral participation \cite{bargsted}.
The use of the Internet has been associated with an increase in social capital \cite{horrigan2001,v2001internet}, leading to a positive effect in political participation. Mainly, the more people are involved in online organizational and political activities, the more they are involved in such activities offline \cite{wellman2001}. Moreover, use of the Internet for information exchange (i.e., searching for information and sending email) has a positive impact on civic engagement\cite{v2001internet} for Generation X (See Supplementary note \ref{SN4}).

\section*{Methods}
\label{sec:data}

The new constitution-making process held in Chile in 2016 is particularly important worldwide because of its high levels and geographic coverage of participation. In the entire participatory process participated more than $200,000$ citizens. Here, we focus on the local encounters (ELAs).

\subsection*{Participants}
We use data on all the self-convoked encounters (ELAs), involving $106,412$ participants. For each ELA, the data includes the name of the municipality in which the encounter took place, the age and gender of participants, the collectively agreed upon constitutional rights, and the argument texts. We run several statistical models to explain the number of held ELAs in each municipality, between April and August 2016. The data on ELAs, used in this study, is publicly available and was previously systematized, and all the argument texts were normalized \cite{fierro2017} (See Supplementary Methods \ref{SMet1}). Finally, we used complementary data on 345 out of the 346 Chilean municipalities, after excluding the Chilean Antarctic territory. 

\subsection*{Demographic cohorts}
\subsubsection*{Generations}

According to Strauss and Howe \cite{StraussHowe}, historical events are associated with recurring generational archetypes. Given that the constitutional process is a political event, we define age cohorts based on the Chilean political background. In 1973 the Chilean military, led by general Augusto Pinochet, staged a \textit{coup d'état} against the socialist government of Salvador Allende. An authoritarian military regime ruled the country until the transition to democracy in 1989. It has been found that younger cohorts are likely less politicized, since they did not experience the military dictatorship nor the political climax that prevailed during the transition to democracy \cite{carlin}. Here we use the three age cohorts proposed by Lindh \textit{et al} :\cite{lindh}

\begin{itemize}
\item  Born in 1955 or before: were adults (over 18 years old) when the breakdown of democracy occurred in 1973. 
\item  Born between 1956 and 1980: became adults during the dictatorship, and experienced the following democratic period in which Pinochet was Commander in Chief of the Army (until 1998);
\item  Born in 1981 and after: became adults after the return to democracy.
\end{itemize}

Figure~\ref{fig:edad}A shows the composition of national population and citizen participation in ELAs, by gender and generational cohorts. A chi-squared test was performed to test whether the distributions of age and gender are statistically equivalents between the national population and citizen participation in ELAs. In either case we can reject the null hypothesis of the test, i.e.,  the number of people by age and gender is independent of the instance (national population or citizen participation) at the $1\%$ significance level. Therefore, the participation distribution does not differ significantly from the population distribution. The contingency tables and chi-squared tests results can be found in the Supplemental Material (Tables~\ref{tab:age_chi} and \ref{tab:sex_chi}).

\subsubsection*{Geography}

On the other hand, the share of population participating in ELAs was substantively different across the country. 
Figure~\ref{fig:edad}B shows citizen participation at the regional level, and the same variability is observed at the municipality level (see Figures~\ref{fig:norte},\ref{fig:centro} and \ref{fig:sur} of the Supplemental Material). 


\vspace{2mm}

\subsection*{Constitutional dimensions}

During the ELA meetings, the participants were asked to identify the most important constitutional concepts in four dimensions: Values and Principles, Rights, Duties, and Institutions. Given the recent social movement in Chile and the increasing demand on social rights, we focused our analysis on the constitutional concepts within the ``Rights'' dimension. The original list of Rights provided by the government included 45 items, and ELAs participants added 14 new ``rights'' in the process, resulting in 59 rights in total. Among the new concepts,``Social rights'', ``Standard of living'' and ``Right to quality public health care'' reaffirm the increasing demands on social rights in Chile, while ``Respect life from conception'' and  ``Right to make one's own decisions about one's life'' arise as specifications of the original ``Right to Life''.  Figure~\ref{fig:edad}C shows the most voted rights at the national level. The full list of rights available to ELAs participants, along with the new concepts that emerged in the deliberation process, can be found in Section~\ref{sec:elasqq} of the Supplemental Material.


\subsection*{Citizen participation}

\subsubsection*{Statistical model}

Using Ordinary Least Squares (OLS) models, we study the effects of socio-demographic and political variables on citizen participation. Given that each ELA represents a collective deliberation instance, we measured citizen participation as the total number of ELAs held in each municipality, instead of the total number of participants. Nevertheless, the main findings of this work do not significantly change when using the number of participants as a dependent variable (see Supplementary Table~\ref{tab:ols_pob}). A histogram of the ELAs number of participants at the national level is also provided in Supplementary Figure~\ref{fig:size_elas}. The regression model specification is the following:

\begin{equation}
    \log (1 + ELA_i) =  \beta  X_i + e_i,\label{eq:ols}
\end{equation}
where $ELA_i$ is the total number of ELAs held in the municipality $i$, $X_i$ is a vector of covariates, and $e_i$ is a random error term. The coefficients of interest are the $\beta$s, which measure the effect of the covariates on citizen participation. Note that we add $1$ unit to the dependent variable. This correction serves to account for zeroes at the logarithmic transformation \cite{McCune2002Analysis}.

\subsubsection*{Independent variables}
Participants did not provide any personal information about income, education, or political inclination. Therefore, the model is estimated at the municipality-level, using aggregated data on socioeconomic and demographic indicators.

\begin{figure}[!h]
\centering
\includegraphics[width=\textwidth]{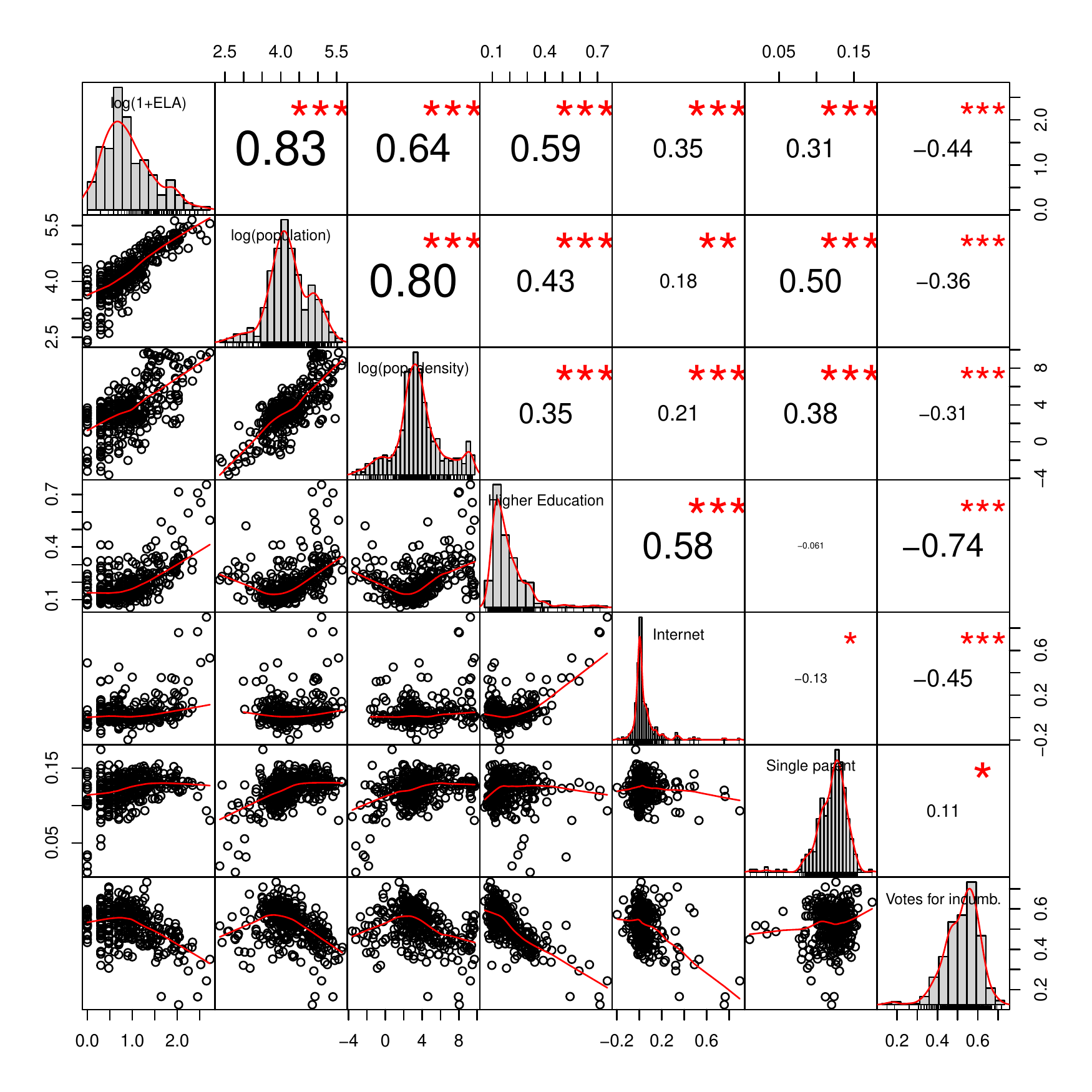}
\caption{Correlogram of main variables.}
\label{fig:correlations}
\end{figure}

We also included political variables to assess the impact of political engagement, such as voter turnout, party affiliation, and political orientation. We further included variables relative to the municipality in order to test whether people were mobilized to participate. We collect all the independent variables from official national sources, such as the population and housing census (CENSO), the Electoral Service (SERVEL), the national municipal information system (SINIM), the national office for regional development (SUBDERE), and the National Socio-Economic Characterization Survey (CASEN). Table~\ref{tab:var_desc} presents a detailed description of the independent variables (see also Supplementary Methods \ref{SMet2}). The descriptive statistics and the correlation for the most relevant variables can be found in the Supplemental Material (Tables~\ref{tab:stats} and Figure~\ref{fig:correlations}, respectively).

\begin{table}[!htbp] \centering 
\scriptsize
  \caption{Variable description and data.} 
  \label{tab:var_desc} 
\begin{tabular}{@{\extracolsep{5pt}}l L{6cm} llll} 
\\[-1.8ex]\hline 
\hline \\[-1.8ex] 
Name & \multicolumn{1}{c}{Description} & \multicolumn{1}{c}{Source} & \multicolumn{1}{c}{Year} & \multicolumn{1}{c}{N} & \multicolumn{1}{c}{Type} \\
\hline \\[-1.8ex] 
Population & Municipal population over 14 years old (only people over the age of 14 were allowed to participate in an ELA). & CENSO & 2017 & 345 & Count \\
\hline
Higher education & Share of municipal population who have successfully completed a higher education degree (advanced technician, bachelor, MSc, PhD.) & CENSO & 2017 & 345 & Continuous (0-1) \\
Internet penetration rate & Constructed as the interaction of two variables from CASEN survey: (i) share of households with at least one internet-connected device, (ii) number of different uses of Internet. & CASEN & 2015 & 324 & Continuous (0-1) \\
SUBDERE groups & Socio-demographic municipal classification based on the dependence on the municipal common fund and the local population. This typology divides municipalities in 7 groups plus an exception group with the highest income municipalities. Group 1 is the most vulnerable, i.e., municipalities in this group have less population and show higher dependency on the municipal common fund. & SUBDERE & 2005 & 335 &  Categorical \\
Poverty & Income poverty rate by municipality, based on income information from CASEN survey, using a method of Small Area Estimation (SAE). & INE & 2017 & 344 & Continuous (0-1) \\
SEDI & Socio-Economic Development Index, which comprises education, income, poverty, housing and sanitation. & OCHISAP & 2013 & 324 & Continuous (0-1) \\  
Community organizations & Number of community organizations in the municipality, such as parent centers, cultural centers, sport clubs, among others. & SINIM & 2015  & 342 &  Count \\
Participation in comm. org. & Share of municipal population that declares to participate in a community organization & CASEN & 2015  & 324 & Continuous (0-1)\\
\hline
Population density & Number of people living in the municipality, per square kilometer (km$^2$)  & SINIM & 2015 & 345 & Continuous\\
Rurality & Share of municipal population living in rural areas. A rural area is defined as an agglomeration with more than 1,000 inhabitants, or between 1,001 and 2,000 inhabitants where more than 50\% of the economically active population is engaged in primary economic activities.  & CENSO & 2017 & 345 & Continuous (0-1) \\
Born in 1981 or after & Share of municipal population born in 1980 or after, which represents the youngest cohort of our study, who became adults after the return to democracy. & CENSO & 2017 & 345 &  Continuous (0-1)\\
Women & Proportion of women in the municipality & CENSO & 2017 & 345 & Continuous (0-1) \\
Single-parent family with children & Share of single-parent families, with children, in the municipality. & CENSO & 2017 &  345 & Continuous (0-1) \\
Two-parent family with children & Share of two-parent families, with children, in the municipality. & CENSO & 2017 & 345 & Continuous (0-1) \\
\hline
Party affiliation & Share of municipal population affiliated to any political party. & SERVEL & 2016 & 345 & Continuous (0-1)\\
Voter turnout & Voter turnout in 2013 presidential elections at the municipal level. & SERVEL & 2013 & 345 & Continuous (0-1)\\
Votes for standing president & Share of votes received by the winning candidate in 2013 presidential elections by municipality. & SERVEL & 2013 & 345 & Continuous (0-1)\\
Municipal officials & Share of municipal population employed by the city hall. & SINIM & 2015 & 345 & Continuous (0-1) \\
Mayor & 3 dummy variables to take into account the political party which supported the winning candidate for mayor, during the 2012  municipal elections:  the first one is equal to 1 if the party is within the government coalition, and 0 otherwise; the second one is equal to one if the party is in the opposition, and 0 otherwise; the third dummy variable is assigned to 1 when the mayor ran for office with no formal party support. & SERVEL & 2012 & 345 & Categorical\\
Incumbent mayor & Dummy variable that takes the value 1 if an incumbent mayor is reelected, and 0 otherwise. & SERVEL & 2012 & 345 & Categorical \\
Government influence & Consists of the sum of: (i) the share of votes obtained by the pro-government deputies, relative to the total votes obtained by both elected deputies; (ii) 1, if the mayor was supported by the government coalition, and 0 otherwise. & SERVEL & 2012 & 345 & Continuous (0-2) \\
\hline \\[-1.8ex] 
Evangelical Christians & Share of municipal population who declared to profess an evangelical Christian religion. & CENSO & 2012 & 341 & Continuous (0-1) \\
\hline \\[-1.8ex] 
\multicolumn{6}{p{16cm}}{Notes: (i) The administrative division of Chile consists of 346 municipalities, from which we excluded the municipality of Antártica because of its special situation. (ii) Sources: Population and housing census (CENSO); Electoral Service (SERVEL); National municipal information system (SINIM);  National office for regional development (SUBDERE); National Socio-Economic Characterization Survey (CASEN); Public Health Observatory in Chile (OCHISAP); National Institute of Statistics (INE). (iii) CASEN survey lacks representativity at municipal level.}
\end{tabular} 
\end{table}

\subsection*{Text analysis}

To analyze all the argument texts of each self-convoked encounter, we use Structural Topic Modeling (STM) \cite{roberts2014stm}, which is a widely-used method for text analyzing \cite{roberts2014, bohr2018, lindstedt2019, almquist2019, herrera2019understanding}. As every topic modeling algorithm, STM assumes that each document in a text corpus is produced from a mixture of latent topics, estimated from the text data, which in turn consists of a collection of words \cite{Blei2012}. Besides, STM enables the use of document-level metadata information, whose goal is to discover latent topics and estimate their relationship to the document metadata. Thus, this method serves as the bridge between statistical techniques and research goals. (See Supplementary Methods \ref{SMet3} for a detailed explanation of Structural Topic Modeling).

\section*{Results}
\label{sec:results}

\subsection*{Citizen participation in ELAs}

Table~\ref{tab:all} reports standardized ordinary least-squares (OLS) regressions of the number of ELAs on several socio-demographic and political variables. The number of ELAs, population, and the number of organizations are count variables, and Mayor, Incumbent Mayor, and SUBDERE groups are categorical. The remaining variables are expressed as the share of the population. For instance, Rurality represents the share of the municipality-level population living in rural areas. Variables with high skewness were transformed with a base 10 log function. The transformed variables are the number of ELAs, population, population density, and the number of community organizations. The p-value of the Ramsey Regression Equation Specification Error Test (RESET) is provided in the caption of each table throughout this work. Only significant interactions are shown in Table~\ref{tab:all}, but the full list of interactions can be found in the Supplemental Material (Table~\ref{tab:SM_full}).

The number of ELAs strongly depends on the population. As expected, this variable is the most variance-explicative predictor of citizen participation. Model 1 Table~\ref{tab:all} includes variables for socio-economic and social capital. This model shows significant and positive effects for higher education, internet penetration rate, and the number of community organizations. Model 2 includes demographic variables, where population density and two-parent family (with children) show significant and negative effects. Finally, Model 3 (for a performance plot see Supplementary Figure \ref{fig_sim}) incorporates political variables.

The share of votes for the incumbent president has a significant and positive effect on the number of ELAs, as well as the percentage of people who are affiliated to any political party. The non-significant effects of Mayor and Government influence (see Supplementary Talble \ref{tab:SM_gov}) variables suggest no evidence of political mobilization by mayors or deputies was found. Finally, the share of Evangelical Christians shows a negative effect that, although small in magnitude, is highly significant. The variance inflation factors are all less than 5 except, for education ($5.5$) and population ($10.3$) (Supplementary Table \ref{tab:SM_VIF}). Considering that the dependent variable in equation \ref{eq:ols} is a count variable, we run a negative binomial regression as an additional robustness check, which accounts for over-dispersion (see Supplementary Table~\ref{tab:SM_poisson}). Comparison with Model 3 Table~\ref{tab:all} shows no significant changes for the main results. Therefore, all subsequent models are estimated using OLS.   

\begin{table}[!h] \centering 
\scriptsize
  \caption{OLS Regressions, p-value RESET test Model 1 = 0.109 , p-value RESET test Model  2 = 0.1418, p-value RESET test Model 3 = 0.3501. RESET test were performed on the second power of regressors. (See a Supplementarty Figure \ref{fig_marginals} for the marginal effects plot)} 
  \label{tab:all} 
\begin{tabular}{@{\extracolsep{-10pt}}lD{.}{.}{-3} D{.}{.}{-3} D{.}{.}{-3} } 
\\[-1.8ex]\hline 
\hline \\[-1.8ex] 
 & \multicolumn{3}{c}{\textit{Outcome variable:}} \\ 
\cline{2-4} 
\\[-1.8ex] & \multicolumn{3}{c}{log (1 + ELAs)} \\ 
\\[-1.8ex] & \multicolumn{1}{c}{(1)} & \multicolumn{1}{c}{(2)} & \multicolumn{1}{c}{(3)}\\ 
\hline \\[-1.8ex] 
  Log (population) & 0.528^{***} & 0.718^{***} & 0.778^{***} \\ 
  & (0.131) & (0.142) & (0.164) \\ 
  \hline \\[-1.8ex]
  Higher Education & 0.156^{***} & 0.128^{**} & 0.185^{***} \\ 
  & (0.047) & (0.050) & (0.065) \\ 
  Internet penetration rate & 0.113^{***} & 0.105^{**} & 0.103^{***} \\ 
  & (0.036) & (0.041) & (0.039) \\ 
  \textit{SUBDERE groups}$^{(1)}$ (control) & \multicolumn{1}{c}{yes}  & \multicolumn{1}{c}{yes} & \multicolumn{1}{c}{yes} \\
  &&& \\
  Log (community organizations) & 0.103^{**} & 0.059 & 0.085^{*} \\ 
  & (0.048) & (0.056) & (0.051) \\ 
  \hline \\[-1.8ex]
  Born in 1981 or after &  & -0.032 & -0.008 \\ 
  &  & (0.052) & (0.056) \\ 
  Rurality &  & -0.081 & -0.052 \\ 
  &  & (0.056) & (0.054) \\ 
  Log (population density) &  & -0.100^{**} & -0.139^{***} \\ 
  &  & (0.050) & (0.053) \\ 
  Women &  & 0.067 & 0.066 \\ 
  &  & (0.072) & (0.069) \\ 
  Two-parent family (with children) &  & -0.133^{***} & -0.093^{**} \\ 
  &  & (0.040) & (0.043) \\ 
  Single-parent family (with children) &  & -0.065 & -0.101^{**} \\ 
  &  & (0.041) & (0.043) \\ 
  \hline \\[-1.8ex]
  Votes for current president &  &  & 0.148^{***} \\ 
  &  &  & (0.046) \\ 
  Municipal officials &  &  & -0.040 \\ 
  &  &  & (0.170) \\ 
  Voter turnout &  &  & 0.087 \\ 
  &  &  & (0.054) \\ 
  Mayor (government) &  &  & 0.063 \\ 
  &  &  & (0.069) \\ 
  Mayor (opposition) &  &  & -0.099 \\ 
  &  &  & (0.079) \\ 
  Party affiliation &  &  & 0.161^{**} \\ 
  &  &  & (0.066) \\ 
  Incumbent Mayor (True) &  &  & -0.020 \\ 
  &  &  & (0.054) \\ 
  Evangelical Christians &  &  & -0.075^{***} \\ 
  &  &  & (0.027) \\ 
  \hline \\[-1.8ex]
  Log(comm. org.) * Log (pop. density)$^{(2)}$ &  & -0.071^{*} & -0.054 \\ 
  &  & (0.041) & (0.041) \\ 
  Constant & -0.072 & 0.042 & 0.037 \\ 
  & (0.192) & (0.203) & (0.209) \\ 
 \hline \\[-1.8ex] 
Observations & \multicolumn{1}{c}{313} & \multicolumn{1}{c}{313} & \multicolumn{1}{c}{310} \\ 
R$^{2}$ & \multicolumn{1}{c}{0.784} & \multicolumn{1}{c}{0.804} & \multicolumn{1}{c}{0.834} \\ 
Adjusted R$^{2}$ & \multicolumn{1}{c}{0.777} & \multicolumn{1}{c}{0.790} & \multicolumn{1}{c}{0.814} \\ 
Residual Std. Error & \multicolumn{1}{c}{0.463 (df = 301)} & \multicolumn{1}{c}{0.449 (df = 291)} & \multicolumn{1}{c}{0.424 (df = 275)} \\ 
F Statistic & \multicolumn{1}{c}{99.586$^{***}$} & \multicolumn{1}{c}{56.800$^{***}$} & \multicolumn{1}{c}{40.652$^{***}$} \\ 
F Statistic & \multicolumn{1}{c}{ (df = 11; 301)} & \multicolumn{1}{c}{ (df = 21; 291)} & \multicolumn{1}{c}{ (df = 34; 275)} \\ 
\hline 
\hline \\[-1.8ex] 
\multicolumn{4}{p{12cm}}{Note: $^{*}$p$<$0.1; $^{**}$p$<$0.05; $^{***}$p$<$0.01} \\
\multicolumn{4}{p{12cm}}{$^{(1)} \quad $ SUBDERE groups are a socio-demographic classification based on county's dependence on the municipal common fund.} \\
\multicolumn{4}{p{12cm}}{$^{(2)} \quad $ Only significant interactions are shown.}
\end{tabular} 
\end{table}

\subsection*{Constitutional rights selection}
\label{sec:ccs}

How do the socio-political determinants of citizen participation impact the selection of constitutional rights? To address this question, we use the most relevant determinants of citizen participation found in the previous section. These are (i) Higher Education; (ii) Population density; and (iii) Votes for current president. We also add other relevant variables, (iv) SEDI (Socio-Economic Development Index); (v) share of municipality population which declare to participate in primary economic activities (CENSO). Finally, given that the Santiago Metropolitan Region exhibits different living dynamics than all other regions of Chile, we included a variable to compare the municipalities belonging to the Santiago Metropolitan Region to all the other municipalities of Chile. We named this last variable as ``Geographic". 

\begin{figure}[!t]
\centering
\includegraphics[width=0.85\textwidth]{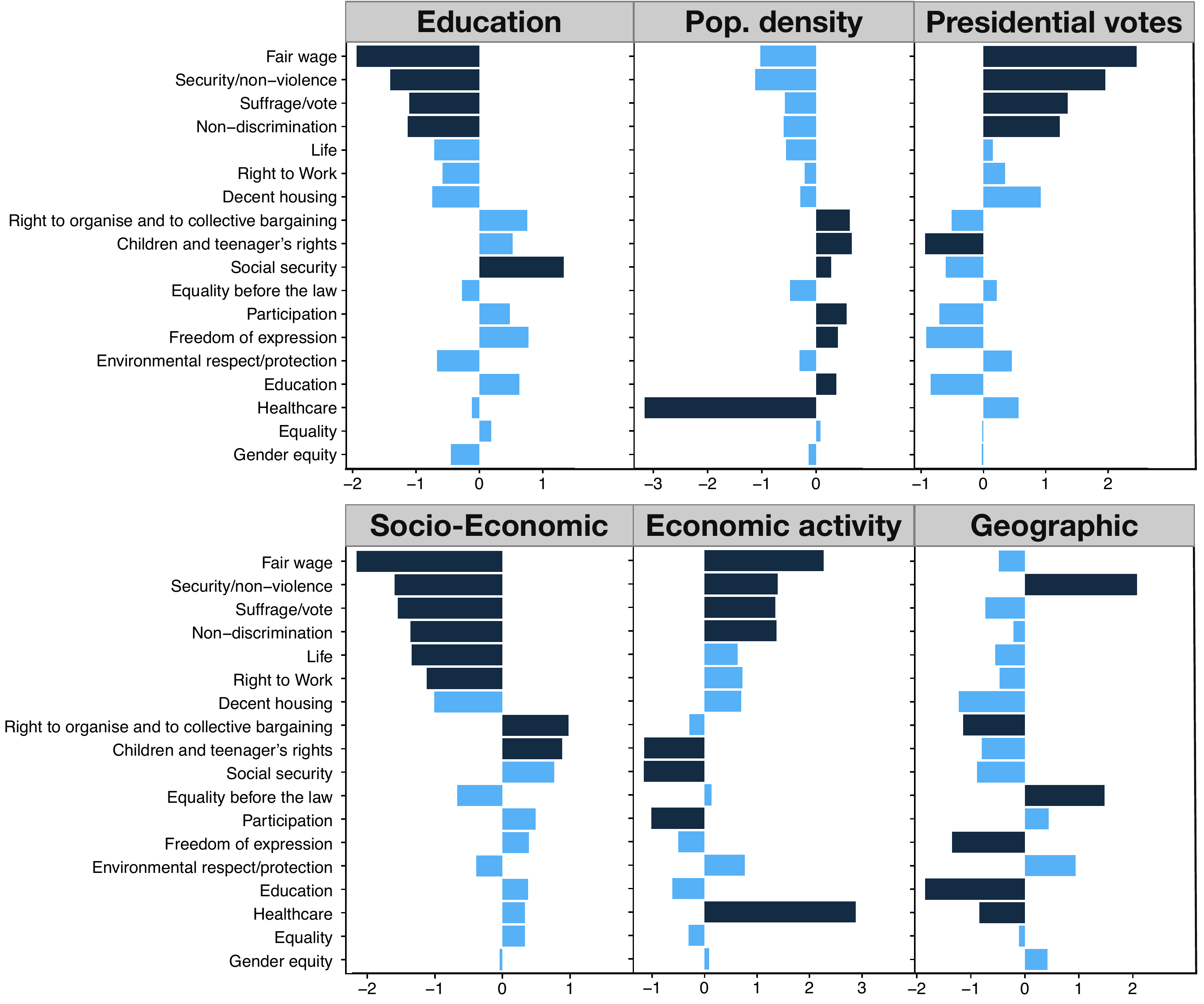}
\caption{What constitutional rights do people select at the municipality level? Differences are shown in the proportion of selection for the 18 most voted concepts. From left to right, municipalities are grouped by the top and bottom quartiles for higher education, population density, votes for the current president, SEDI (Socio-economic Development Index), primary economic activity, and geographical zone.  For each citizen participation determinant, the bar length is calculated as the difference between the proportion of mentions of a constitutional right in municipalities in the top quartile and its proportion in municipalities in the bottom quartile. For instance, a positive (negative) value means that the more (less) educated municipalities select that concept in a higher proportion than less (more) educated counties. Particularly, the constitutional right to social security is selected significantly more in municipalities whose share of people with higher education is in the top quartile in the country. In simple words, the most educated municipalities select significantly more the right to social security, when compared to the less educated municipalities. Bar color: dark blue when the $\chi ^2$ test is significant at $99\%$, and light blue otherwise.} 
\label{fig:derechos}
\end{figure}

For each variable, except Geographic, we consider the municipalities in the top and bottom quartiles. Then, we compared the number of times each constitutional concept was selected, as a proportion of the total number of constitutional concepts selected in those municipalities. This comparison was performed using a 2-sample test for equality of proportions ($\chi ^2$ test). For example, the first panel of Figure~\ref{fig:derechos} shows the difference in concept selection between the top 25 percent and the bottom 25 percent of municipalities for Higher Education. For each concept, the bar length is calculated as the share of that concept in the top $25\%$ group, minus the share of that concept in the bottom $25\%$ group. Therefore, a positive (negative) value means that the more (less) educated municipalities selected that concept in a higher proportion than less (more) educated counties. For the Geographic variable, the municipality separation was made based on the region to which they belong. A positive value means that the municipalities belong to the Metropolitan Region selected those concepts in higher proportions.

\subsection*{Topic modeling}

The argument texts were analyzed using the Structural Topic Model (STM), a tool that enables estimating topics including document-level metadata. The corpus consisted of all argument texts that were written for constitutional Rights throughout all ELAs. Each text corresponds to a document of the corpus, and the constitutional concept (to which the text refers) was incorporated as prevalence document metadata.

\begin{table}[h!]
\scriptsize
  \caption{OLS regressions results for STM. Table shows the top three categories for each regression. Concepts in italic font were not included in the original list of concepts proposed by the government, and were added by ELAs participants. The last two columns in the table show the word sets for three topics. The words shown here have been translated into English. All the STM analysis has been performed with the original texts in Spanish.} 
  \label{tab:coef_stm3} 
\begin{center}
\begin{tabular}{L{2cm} L{4cm} l L{3cm}  L{4cm}}
\\[-1.8ex]\hline 
\hline \\[-1.8ex] 
 Topic & Right & \textit{Outcome variable:} & Highest Probability & Frequent and exclusive\\ 
\cline{3-3} 
 \\[-1.8ex] 
 & & \multicolumn{1}{c}{Topic} &  &  \\ 
\hline \\[-1.8ex] 
Education & Education        & $0.336 \; (0.004)^{***}$ & \multirow{9}{3cm}{education, quality, for-free, must, access, universal, public, level, free education, (there must) be education, profit, public education, secular, integral, free-of-charge, public for-free, higher, opportunity, civic} & \multirow{9}{4cm}{for-free secular, integral education, quality for-free, university, teacher, room, public free education, free education, free-of-charge, public education, higher education, (there must) be education, must guarantee education, decent education, (there must) be free education, guarantee education, student, free access} \\
& \textit{Right to quality public health care}                 & $0.092 \; (0.014)^{***}$ && \\
& Freedom of Education                               & $0.091 \; (0.007)^{***}$ && \\
&&&&\\
&&&&\\
             &&&&\\
             &&&&\\
             &&&&\\
             &&&&\\
             &&&&\\
\hline \\[-1.8ex] 
Equality  & Equality before the law        & $0.345 \; (0.005)^{***}$ & \multirow{9}{3cm}{equality, law, must, same, to-exist, justice, (there must) be equality, opportunity, must exist, treatment, access, same right, process, privilege, gender, to-treat, can, egalitarian} & \multirow{9}{4cm}{must exist difference, same condition, due process, have equality, military, exist difference, (there must) be equality , justice, same right, equal treatment, equality, law, to-exist equality, privilege, must exist equality, same treatment, to-exist privilege, same opportunity, must have equality} \\
& Access to justice / due process                 & $0.286 \; (0.009)^{***}$ && \\
& Equality                              & $0.258 \; (0.005)^{***}$ && \\
&&&&\\
&&&&\\
             &&&&\\
             &&&&\\
             &&&&\\
             &&&&\\                     
\hline \\[-1.8ex] 
Security & Security / non-violence                       & $0.358 \; (0.005)^{***}$ & \multirow{8}{3cm}{ to-live, must, violence, security, safe, can, peace, space, to-feel, quiet, get-better, can live, crime, home, house, fear, 	tranquility, (there must) be security} & \multirow{8}{4cm}{(there must) be greater, to-feel, must live, insecurity, neighborhood, crime, (there must) be protection, peace, violence, can live, (there must) be security, quiet, street, safe, tranquility, to-live quietly } \\
& Freedom of movement                               & $0.112 \; (0.019)^{***}$ &&\\
& Decent housing                                        & $0.089 \; (0.004)^{***}$ &&\\
             &&&&\\
             &&&&\\
             &&&&\\
                          &&&&\\
             &&&&\\
             \hline \\[-1.8ex] 
Environment & Environmental respect / protection          & $0.468 \; (0.008)^{***}$  & \multirow{8}{3cm}{must, environment, resource, good, natural, better, nature, natural resource, to-live, water, generation, pollution, free, use, future, to-preserve, respect,	sustainable} & \multirow{8}{4cm}{ecosystem, environment free, better quality, pollution, future generation, healthy environment, environment, better, water, good, must preserve, clean, natural resource, nature, sustainable, sustainability, sustainable development}\\
             & \textit{Right to water}                            & $0.295 \; (0.023)^{***}$ & & \\
             & \textit{Conservation of cultural and historical heritage} & $0.147 \; (0.055)$ & & \\
             &&&&\\
             &&&&\\
             &&&&\\
             &&&&\\
             &&&&\\
\hline \\[-1.8ex] 
\multicolumn{5}{l}{\scriptsize{$^{***}p<0.001$, $^{**}p<0.01$, $^*p<0.05$}} \\
\end{tabular}
\end{center}
\end{table}

\begin{figure}[t!]
    \centering
    \includegraphics[width=1\textwidth]{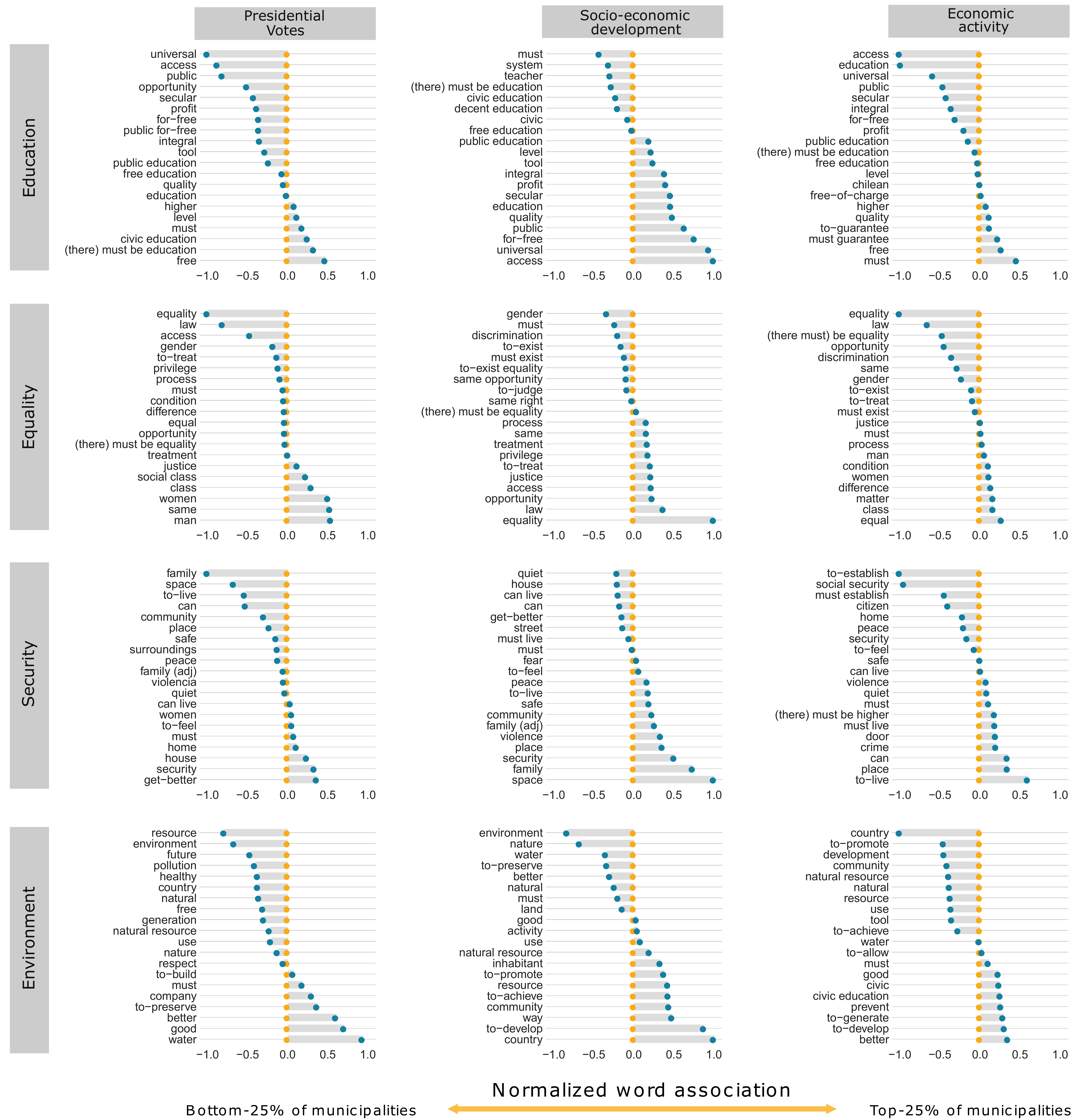}
    \caption{A word comparison of the constitutional rights debate, at the municipality level. We show the emergent topics: Education, Equality, Security, and Environment for three different citizen participation determinants, presidential votes, socio-economic development, and economic activity. Words are oriented along the X-axis based on how much they are associated to the inspected determinant. For instance, in the Education topic and Presidential Votes determinant, the word ``universal,'' which is on the left side, is associated with the municipalities where the elected president got the least of the votes. The word ``free,'' which is on the right side, is associated with the municipalities where the elected president got most of the votes. We note that for the topic Equality,  the word ``process'' comes from ``due process'' and ``treatment'' refers to ``behaviour towards''. Likewise, for the topic Security, ``door'' comes from ``revolving door'', which refers to inmate release and recidivism, and  ``can live'' comes from ``can live peacefully in our neighborhood''. }
    \label{fig:fit2}
\end{figure}

We estimated an STM with 23 topics in our corpus. We chose the number of topics by performing various model diagnostics such as Held-Out Likelihood and the Semantic Coherence for a different number of topics (See Supplementary Figure~\ref{fig:ntopic}). We also included the constitutional ``rights" concepts as prevalence covariates in the model. The authors manually assigned the topic labels by looking at two sets of words for each topic. The first is a set of highest probability words, inferred directly from topic-word distribution. The second is a set of words that are both frequent and exclusive for each topic \cite{bischof2012}. 

Table~\ref{tab:coef_stm3} shows both sets of words, for four different topics (Education, Equality, Security, and Environment). These topics were selected as examples because they cover a broad range of themes and show a considerable difference in content when comparing groups across different determinants. The sets of words for the 23 topics can be found in Figures~\ref{fig:wordcloud1} and \ref{fig:wordcloud055} of Supplemental Material, in the original language (Spanish).

We asked how each constitutional concept affects the topic proportion? We ran a regression model in which each document is an observation; the outcome variable is the probability that a specific topic generates each document in the STM model (remember that we model each document as a weighted combination of topics). The explanatory variables are all constitutional rights. Therefore, the estimated coefficients represent, for each topic, the effect of the constitutional rights on the topic.  Table~\ref{tab:coef_stm3} shows the regression results for the same four topics. Since each topic has 59 explanation variables (the total number of constitutional rights); we display here only the top three categories, i.e., the constitutional concepts with the highest topic proportion within each topic, for each regression. The results show agreement between the topics emerging from the argument texts and the concepts that originated those texts. The regression results for the remaining topics can be found in Table~\ref{tab:coef_stm}.

We observe that socio-demographic and political variables alter the frequency in which a particular concept is mentioned (Figure~\ref{fig:derechos}), and it is also likely that those variables affect how topics are discussed. Therefore, we estimate a second STM model with prevalence and topic content covariates. Recalling that, while topic prevalence captures how much each topic contributes to a document, topic content variables enable the vocabulary used to discuss a particular topic to vary. Here, we use three different determinants as covariates: presidential votes, socio-economic development, and primary economic activity. Figure~\ref{fig:fit2} shows the differences in vocabulary by association level with the determinants for four topics: Education, Equality, Security, and, Environment. The list of words for each group and topic, along with representative phrases can be found in Tables~\ref{tab:stm_env} to \ref{tab:stm_eq}.


\section*{Discussion}
\label{sec:disc}


In Table \ref{tab:all} we observe that population density has a negative and significant effect on the number of ELAs that were held in a municipality. This finding is in agreement with previous research reporting that high population density has a negative impact on social capital \cite{mccull}.

Regression results show a positive effect on the share of women and a negative effect on the youngest age cohort. Even though both effects are non-significant, their signs are in agreement with voting behavior in Chile, where the younger cohorts exhibit the lowest participation rate, and voter turnout rates for women exceed those for men \cite{pnud}. 
Our model also indicates that fewer ELAs were organized in municipalities with a higher proportion of single-parent families and two-parent families with children. An intuitive explanation is that people with children have less time to engage in social activities. However, previous research has reported an indirect effect of having children in political participation, through its consequences for female labor force participation and the consequent impact on social capital \cite{schlozman1999}.


According to the literature on political participation and social capital, education has a high impact both on social participation \cite{huang}, and in the understanding of complex political information \cite{verba}. As expected, our results confirm a significant and positive return to education, measured as the share of people with higher education, on the number of ELAs.  Furthermore, from all the significant regressors of the full model, education has the second-highest coefficient, after population (Table \ref{tab:all}). Besides, we include the number of community organizations in our model as a proxy of social capital at the municipality level. The result shows a positive and slightly significant effect after controlling for. However, the coefficient is not robust after controlling for socio-political confounders, suggesting that this variable can be correlated to one of those confounder or that the number of community organizations does not represent actual participation in those organizations.

The socio-economic factor, represented here by the SUBDERE groups, shows no significant effect in the number of ELAs. SUBDERE classification reflects the municipality dependence on the Municipal Common Fund, and it is used here as a proxy on municipal residents' income.  Since the socio-demographic groups created by SUBDERE incorporates the population in the municipality classification, this variable may be correlated with population density. Therefore, Model 3 in Table~\ref{tab:all} (henceforth called "the full model") was repeated, replacing SUBDERE classification with a poverty measure, which also shows no significant effect (Supplementary Table~\ref{tab:SM_pobr}). 
The regression results using poverty are shown in Supplementary Table~\ref{tab:SM_pobr}. Replacing SUBDERE with poverty in the model does not substantially change the effect of the other socio-economic or demographic variables.


Internet penetration rate shows a consistent, positive, and significant effect on the number of ELAs. This result is consistent with previous research showing a positive impact of the Internet on social relations and civic engagement. The result of this comparison is shown in Table~\ref{tab:SM_casen}.  Given this variable has been derived from CASEN survey, we performed three comparisons (Supplementary Method \ref{SMet2}), we show that at considering only the 139 counties where the CASEN survey is representative, the effect of the Internet penetration rate becomes non-significant. However, in the bootstrapped regression the effect is also non-significant and the error increases, suggesting that the loss of statistical significance may be due to the smaller sample.


Regarding political variables, the share of votes going to the winning candidate in the first round of the presidential election has a positive and significant effect on participation (Table~\ref{tab:all}). This effect remains when considering the voting of the second ballot (Supplementary Table~\ref{tab:SM_votos}). This result suggests that people who supported the winning candidate were more willing to organize and participate in an ELA since the whole idea of the new constitution was devised and started by the government they chose. 

On the other hand, voter turnout for the first round of the 2013 presidential election has a positive but non-significant effect in the number of ELAs (Table~\ref{tab:all}). For the runoff election, however, the effect becomes positive and significant (Supplementary Table~\ref{tab:SM_votos}). Note that voter turnout for the second round was about seven percentage points lower than for the first round ($50\%$ in the first round, $43\%$ in the runoff ballot), which is not unusual. The phenomenon of turnout decline in runoff elections has been studied for the past 40 years. In particular, the decline is higher when the election result seemed predetermined, and when the vote proportion received in the first round by candidates not qualifying for the runoff goes up \cite{pierce1981,wright}. The latter implies that the runoff election does not involve the diversity of ideas that led many voters to turn out at the first ballot. In simple words, many of those who abstain from the second ballot do so because no candidate represents them. Therefore, the positive and significant effect of voter turnout in the runoff election may also reflect citizen support for the winning candidate.

Party affiliation is also a positive and significant variable (Table~\ref{tab:all}), although this variable includes political parties both within the government coalition and within the opposition coalition. When the model incorporates only parties within the government coalition or only parties within the opposition coalition, the effect remains positive and significant. This result suggests that municipalities with people more actively engaged in politics have a better disposition to participate in an ELA, even if they dislike the government proposing the constitution-making idea.

Political mobilization was explored by evaluating the political affiliation of the mayor. No significant effect was found. To test whether mobilization was driven by members of Parliament, a new variable was created, which we called ``Government influence`'' and combines the political affiliation of mayors and deputies. In Supplementary Table~\ref{tab:SM_gov}, the variables Mayor and Incumbent Mayor were replaced with this variable, but this effect was not significant.

Finally, the negative and significant effect of the share of Evangelical Christians on the number of ELAs presents a question on the role of religion in society. It has been argued that Churches --in the USA context-- play an important role in building up the civic skills of those least likely to participate in politics \cite{jones2001}. However, this might not be true for Chilean evangelicals who often perceive themselves as second-class citizens, legally and socially disadvantaged in comparison to the members of the Catholic Church \cite{boas2016} and, therefore, might be inclined to disengage from political activity.

\begin{figure}[!h]
\centering
\includegraphics[width=\textwidth]{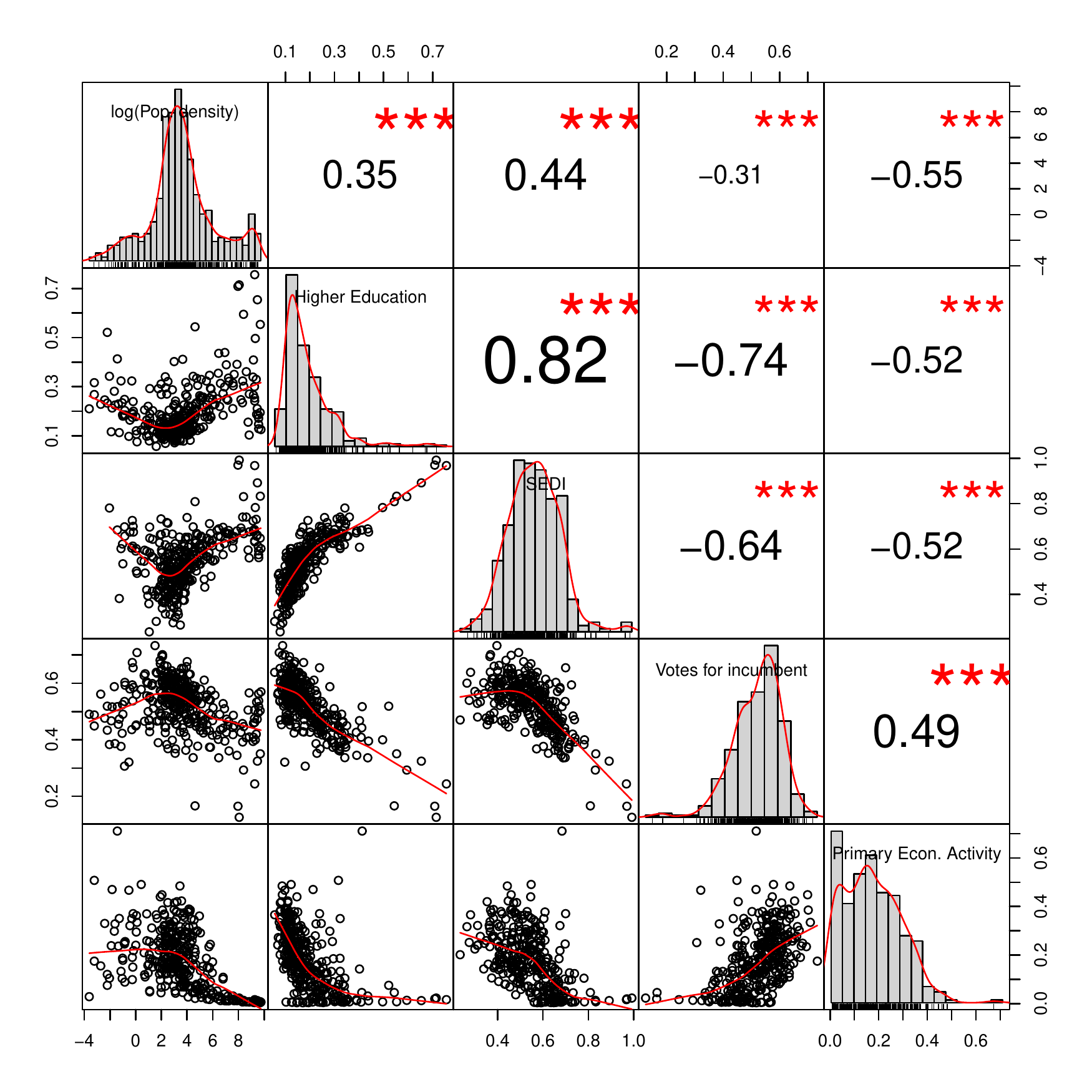}
\caption{Correlogram of variables included in concept selection analysis.}
\label{fig:corr_short}
\end{figure}

A significant variation on constitutional concept selection (Figure~\ref{fig:derechos}) suggests that the previously found socio-political determinants of citizen participation also impact the selection of constitutional rights. At first, let us note that \textit{Fair Wage}, \textit{Security/Non violence} and \textit{Non-discrimination} show a negative and significant variation for ``Higher education'' and SEDI, and a positive and significant variation for ``Votes for current president'' and ``Primary economic activity''. These effects may reflect the needs of a vulnerable population, subject to precarious working conditions and a violent environment, for better and safer living conditions. Also, the share of votes for the current president serves as a proxy for left-wing political orientation, which is usually characterized by the concern for social welfare and equality. The co-occurrence of significant differences in socioeconomic variables (and their signs) are consistent with the correlation among those variables, i.e., ``Higher education'' correlates positively with SEDI and negatively with ``Votes for current president'' and ``Primary economic activity'' (Supplementary Figure \ref{fig:corr_short}).

Regarding \textit{Education} and \textit{Healthcare}, there is a significant difference between the municipalities within the Metropolitan Region and the municipalities in the rest of Chile (panel ``Geographic'' of Figure~\ref{fig:derechos}). As expected, since Chile is one of Latin America's most highly centralized countries, the latter group selected both constitutional rights more frequently, which points to a lack of access to health and educational centers. This result is particularly crucial for \textit{Healthcare}, which also shows a significant difference for ``Population density'' and ``Primary economic activity''.  Finally, the right to \textit{Security/Non-violence} was highlighted by counties within the metropolitan region, which suggest a particular concern about crime in this zone.

\begin{figure}[t!]
    \centering
    \includegraphics[width=1\textwidth]{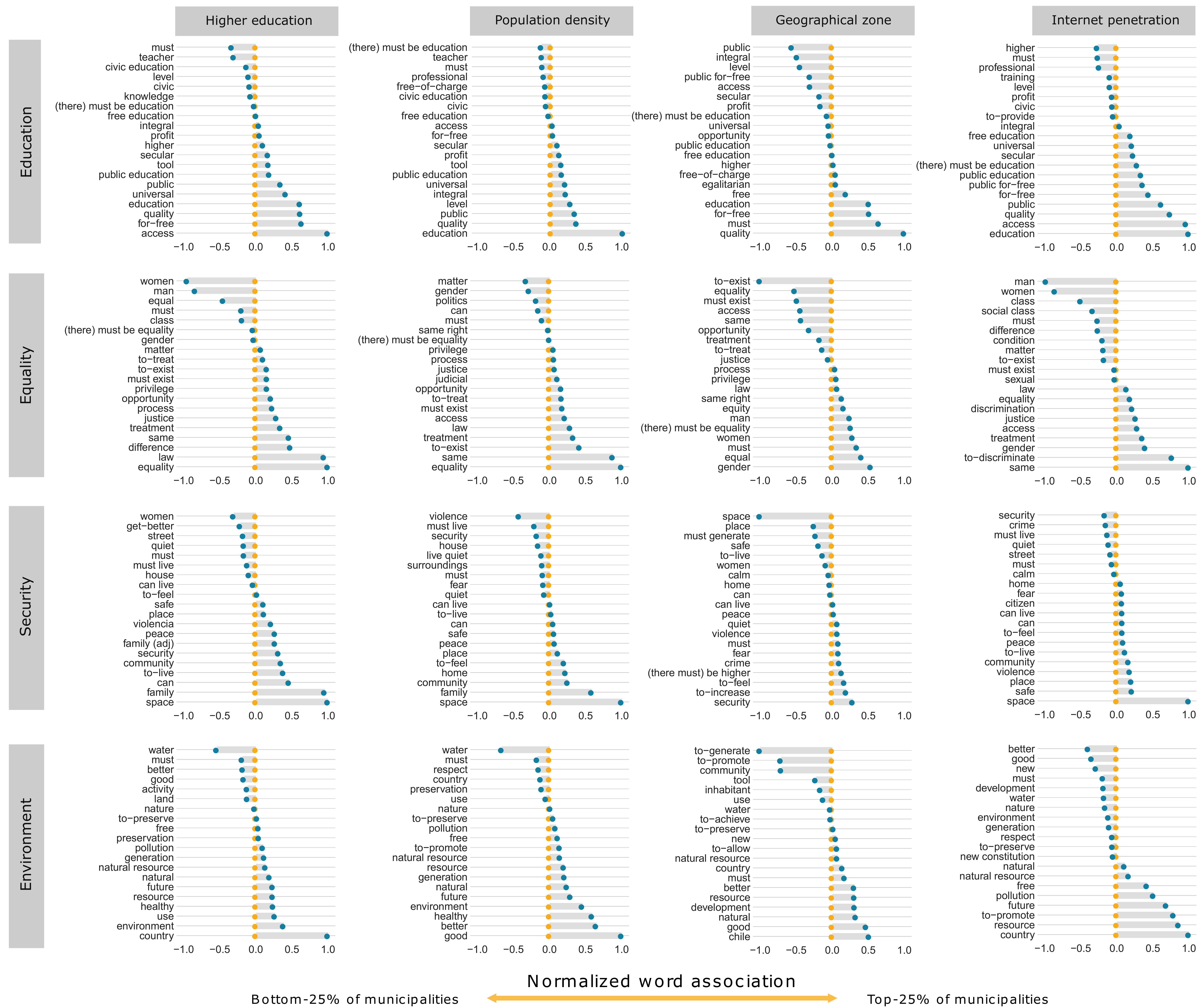}
    \caption{A word comparison of the constitutional rights debate, at the municipality level. We show the emergent topics: Education, Equality, Security, and Environment for four different citizen participation determinants. Words are oriented along the X-axis based on how much they are associated to the inspected determinant. We note that for the topic Equality,  the word ``process'' comes from ``due process'' and ``treatment'' refers to ``behaviour towards''. Likewise, for the topic Security, ``door'' comes from ``revolving door'', which refers to inmate release and recidivism.}
    \label{fig:fit2SM}
\end{figure}

Figure~\ref{fig:fit2} shows how the share of votes for the current president, which reflects support for the government, as well as other socio-demographic variables, changes the way topics are discussed. In the case of the topic ``Environment'', the word ``water'' is strongly associated with the top government supporting municipalities (left-wing), and with municipalities with low population density, and a lower proportion of highly educated people (Supplementary Figure \ref{fig:fit2SM}). This is also related to the concept \textit{Right to water}, that emerged during the ELAs, and claims the nationalization of natural resources. On the other hand, the topic ``Education'' shows a transverse support for free and public education, while words like  ``access'' and ``universal'' predominates in highly educated, more highly socioeconomically developed, and right-wing municipalities. For topic ``Equality,'' highly educated, more socioeconomic developed, and right-wing municipalities discuss ``law'' and ``equality'' more, while their opposite groups are more concerned about gender equality and social classes.

Topic modeling also enables us to explore the emergent content of the whole corpus. Given that each document is associated with a particular constitutional concept, the STM shows us how different constitutional concepts mix into topics. For instance, the topic ``Equality'' (Table~\ref{tab:coef_stm3}) contains three related concepts of similar proportions (\textit{Equality before the law, Access to justice}, and \textit{Equality}).
This mixture of topics is especially relevant in this context, for many of the concepts were added by the participants, although they were variations or combinations of existing concepts.  For example, the topic ``Environment'' comprises the \textit{Right to water}, which was not included in the original list. Figure~\ref{fig:fit2} shows that the word ``water'', probably arising from the  \textit{Right to water},  is strongly associated with a political position. Finally, it is worth noting that the concept \textit{Freedom of movement} is associated to the topic ``Security'', suggesting that this right is interpreted as ``moving safely'' rather than the right of  travel from place to place, or changing the place where one resides or works.

Words have different psychological properties, and tend to be processed in the brain very differently \cite{miller1995, tausczik2010}. Therefore, the terms shown in Figure~\ref{fig:fit2} can also be categorized according to their linguistic usage. Typically, a content/style classification is used to distinguish between \textit{what} people are saying, i.e., the content of a communication, and \textit{how} people are communicating, the style words. Since most of the style words, such as nouns, many adjectives and adverbs, have been removed from the corpus for the purpose of the topic modeling, here we classify between nouns/adjectives and verbs/verbal phrases. 

At first, we classified each term as ``action'' or ``concept''. For example, all terms containing verbs, such as ``must'', ``can live'' or ``(there must) be education'',  are classified as actions, while nouns like ``security'' and ``profit'', and adjectives like ``safe'' and ``familiar'', are classified as concepts. In general, all topics show more concepts than actions. However, by looking at  Figure~\ref{fig:fit2}, and counting actions and concepts across the two levels in all variables, some comparisons can be drawn. For instance, municipalities with a higher socio-economic development index use more concepts than their less developed peers, particularly for topics ``Equality'' and ``Security''. On the other hand, for topic ``Security'', the municipalities with a higher share of people working in primary economic activity use more actions in their arguments, in comparison to the opposite group. It is therefore hypothesized that conceptualization is related to knowledge. Knowledge facilitates proposing themes, concepts and ideas, so that the greater the knowledge, the greater the conceptualization. 

Furthermore, the use of verbal forms such as ``There must be...'', ``must have'', ``can live'' are linked to normative or prescriptive intentions. In many of our cases, the terms classified as ``actions'' are actually this type of verbal form. Let us take a look at the topic ``Security'', for the socio-economic development variable (Figure~\ref{fig:fit2}). The four actions associated with the less developed municipalities are ``can live'' , ``can'', ``must live'', ``must''.  This would indicate a predominantly normative intention from this group.  However,  a sentiment analysis also reveals that the terms used by them score higher (positive) in comparison to the more developed municipalities (Supplementary Table~\ref{tab:stm_sent}). This is mainly caused  by the predominant use of verbs by this group. In general, verbs in this corpus have a positive connotation: ``get-better'', ``guarantee'', ``can live''. All this suggests that this normative intention may be driven by the sentiment: If the matter concern you, you probably won't produce conceptually dense arguments; rather, you will express a desire, a proposal for action.

\section*{Conclusions}

The Chilean constitution-making process of 2016 was a unique experiment in terms of the political history of the country, and its unprecedented high level of participation and territory coverage compared to similar processes held in other countries \cite{ocde}. Taken together, our results show that engagement in politics and support for the government increased participation. However, no evidence of political mobilization by mayors or deputies was found, which suggests that citizen involvement in the constitutional process was ideologically, but voluntarily biased. As the OECD report conjectures, ``those citizens who support the acting government may be more likely to participate in the consultation, even when all citizens are given that opportunity''\cite{ocde}. On the other hand, the ranking of constitutional concepts, specifically of constitutional rights, shows different priorities for different groups, which may result from social needs or political beliefs. Finally, using a structural topic modeling approach, we find that the emerging content from ELAs can be mapped ideologically. Moreover, We find that people in more knowledgeable municipalities, i.e., high levels of education,socio-economic development, and complex economic structure, use more themes concepts, and ideas than actions, suggesting that knowledge facilitates proposing themes, concepts, and ideas, so that the greater the knowledge, the greater the conceptualization. These results can inform new deliberative and massive consulting processes, which in the near future may be a common means of political participation.




\section*{Data Availability}
The necessary data at municipality level to reproduce this work can be delivered under a reasonable request to the authors. The data on the constitutional process can be found here:\\ \url{http://datos.gob.cl/dataset/proceso-constituyente-abierto-a-la-ciudadania}.

\section*{Code availability}
The entire analysis and data processing were done using the standard R 3.6.3 libraries (https://www.r-project.org/). For Structural Topic Modeling we used the standard \textit{stm} R package developed by Roberts, Stewart, and Tingley (2018) \cite{stm}.

\section*{Acknowledgments}
The authors acknowledge the financial support of the Centro de Invetigación en Complejidad Social (CICS) at Universidad del Desarrollo, and to Dr. Anne Bliss for the helpful insights and discussions.

\section*{Author Contributions}
M.P.R. contributed to the data cleaning and analysis, interpretation of data, and writing the manuscript\\
C.C., contributed to the study conception and design, interpretation of data, and writing the manuscript.\\
J.P.C, C.F., and C.R-S. contributed to the interpretation of data and proof-reading the manuscript.\\
C.F. contributed to the data cleaning, text normalization, and data interpretation.

\section*{Corresponding Authors}
M. P. Raveau (mraveaum@udd.cl).\\
Cristian Candia (cristian.candia@kellogg.northwestern.edu | ccandiav@mit.edu | ccandiav@udd.cl).

\bibliographystyle{unsrt}
\bibliography{bibrav}

\pagebreak
\clearpage

\setcounter{page}{1}
\beginsupplement

\section*{Supplementary material}

\section{Supplementary Notes}

\subsection{Socio-demographic factors and political participation}\label{SN1}

In Latin America, the gender gap index by the World Economic Forum shows a low performance in political empowerment, compared to economic participation and opportunity, educational attainment, and health \cite{pachon2012}. Latin American women are less likely to be involved in political activities, especially in  protest demonstration and party affiliation. This phenomenon is closely related to women's occupations; in general, the greater the female labor force participation, the greater their political participation \cite{pachon2012}. A 1999 study on gender, employment, and political participation found a number of factors that are fostered at work and enhance political activity, such as getting requests for activity on the job, supervising others, and exercising civic skills \cite{schlozman1999}. For women, the experience of being discriminated against on the basis of sex would also lead to political activity. Furthermore, being married and having children at home, indirectly depresses political activity through consequences for workforce participation \cite{schlozman1999}. However, a recent study on electoral turnout in Chile shows that for almost all age groups voter turnout rates for women exceeds voter turnout rates for men \cite{pnud}.

The relationship between voter turnout and age is well-established in the literature. The level of participation is relatively low during early adult life, then gradually increases among middle-aged voters, and softly declines with old age \cite{wolfinger1980}. This behavior has been connected to a variety of adult-roles, such as settling down, marriage, home ownership, getting a job, and leaving school \cite{highton2001}. However, in Chile  there is also a generational effect in voter turnout due to the electoral system. Those who registered to vote in the 1988 plebiscite were enrolled in a system with mandatory voting until 2011, which would explain the increase in electoral participation with age. This behavior is not directly associated with the age of people, but with their registration rate. Those who reached the voting age after the plebiscite show lower enrollment and participation rates \cite{contreras}.

Regarding the socioeconomic resources, income and education level have a positive and significant effect on political participation. Economic and cultural resources provide the intellectual and cognitive skills that typically reduce the subjective costs of participation \cite{verba}. In Chile, the young electorate is strongly class-biased, particularly by income, which causes voter registration rates in the upper class to double those of lower class citizens \cite{corvalan}. At the municipality level, an analysis of the 2016 municipal election in Chile showed a negative effect of income on voter turnout \cite{pnud}. However, this effect reverses when the observations are contained to Santiago Metropolitan Region. This implies that spacial socioeconomic segregation is stronger in Santiago than in the rest of Chile, where different socioeconomic classes coexist in the same municipality \cite{pnud}. 


\subsection{Social capital in political participation}\label{SN4}

Since ELAs are group activities, social capital emerges as a relevant predictor of citizen participation. The term ``social capital'' was introduced by Coleman (1988) as a conceptual tool to describe how rational or purposive action is shaped by the social context \cite{coleman1988}. Just as physical capital improves tools to facilitates production, and human capital brings people new skills and capabilities, social capital relies on the relations among persons that facilitate action. There is empirical evidence suggesting that the production of social capital encourages citizens to become more engaged in a broader range of traditional political activities \cite{ladua}. In Latin America this relation holds true: a study conducted using data from Argentina, Chile, Mexico and Peru, shows that greater involvement in non-political organization, as well as higher levels of interpersonal trust, leads to more participation in political activities \cite{klesner}. 

In a municipality-wide study of electoral participation in Chile, Bargsted \textit{et al} \cite{bargsted} use two proxy variables to measure social capital: population density and rural population. High population density would have a negative impact in social capital and social organization for men \cite{mccull}, which would negatively affect voter turnout. On the other hand, studies on the role of urban and rural residency in promoting political activity have reached mixed conclusions. While some studies had found that urbanization would make political participation easier,\cite{campbell,muller1970} others have shown that isolation and the lesser availability of public service in rural areas increase the sense of responsibility to others \cite{hofferth}, promoting social capital and therefore civic engagement. The evidence in Chile shows a significant negative effect of population density on electoral participation, but no significant effect of rural population \cite{bargsted}.

On the other hand, participation in voluntary association is commonly used to measure social capital \cite{putnam1995}.  This idea is based on the assumption that membership in voluntary associations generates trust and facilitates cooperation among members. Also, citizens that join voluntary organizations usually meet more people,  expand their social circle, and hence become more engaged in civic life \cite{verba}.  However, participation in associations represent only a small part of human interaction. Other networks, such as  family, schools, work, media and internet, have a strong impact on norms and values \cite{westlund}. Therefore, the role of associations in social capital generation may be limited. Recent research has addressed the role of media in the production of social capital. The evidence suggests that  use of the Internet supplements network capital by extending existing levels of personal and telephone contact, therefore increasing social contact and civic engagement \cite{horrigan2001,v2001internet}. Thus, the more people are involved in online organizational and political activities, the more they are involved in such activities offline \cite{wellman2001}.

As well as the electoral participation, social capital is influenced by socio-demographic factors, such as education, age and gender. Education plays an important role in social capital production. Schooling not only spreads knowledge, but also cultivates social norms, which is the core of social capital. In schools, students learn the basic norms and responsibilities in society, and values such as reciprocity, respect and trust. A meta-analysis of the effect of education on social capital shows a significant and positive return to education on individual social capital \cite{huang}.  It has been found that social capital, measured as occupational network resources, also varies across the life course. While daily social interaction decreases with age, social resources from occupational networks and voluntary memberships tend to increase with age, leveling off among older respondents \cite{mcdonald}. Regarding gender, being male as opposed to being female appears to increase the probability of group membership \cite{greek}.  Finally, marital status and the presence of children at home have also been considered as control variables in social capital and electoral participation literature \cite{soroka,greek,carlin}.


\pagebreak

\section{Supplementary methods}

\subsection{Argument texts}\label{SMet1}
Along with the constitutional concept selection, each ELA had to write, for each selected concept, a short argument about why this concept is relevant and should be included in the new constitution. These texts were subsequently processed and classified by Fierro $et$ $al$ (2017) \cite{fierro2017}. At first, they classified all the arguments using three kinds of propositions: facts (what it “is”, “was”, or “will be”), values (propositions representing evaluation statements), and policies (“what should be done?”). Then, each argument was normalized to follow the structural pattern subject – verb – direct object -  complement. In this process, more than 120 annotators participated, selected from local undergrad students and professionals in sociology, psychology, political science, linguistics, etc. The classification process was validated by sampling a random set of annotations, which were corrected again by the team.  The estimated error was less than $15\%$. Afterwards they searched for automatic classification tasks that mimic the human processes, testing standard linear classifiers, and simple neural-network based methods. They found that some of the manual tasks are suitable for automatization, particularly the classification of arguments, for which they achieved a $90\%$ top-5 accuracy in a multi-class classification of arguments \cite{fierro2017}.


\subsection{Age and gender}\label{sec:SMdata}

\begin{table}[h!]
\caption{Contingency table for  participants in ELAs and national population, by age cohorts. Pearson's Chi-squared test results for age cohorts: X-squared = 78.989,  p-value $<$ 2.2$\cdot 10^{-6}$}
\label{tab:age_chi}
\centering
\begin{tabular}{|l|c|c|c|}
\hline 
 & Born in 1980   & Born between  & Born in 1955 \\ 
 & or after  & 1956 and 1980 &   or before \\ 
\hline 
National population & 6146410 & 5656007 & 2483653 \\ 
Participants in ELAs & 46078 & 39902 & 18216 \\ 
\hline 
\end{tabular}
\end{table}

\begin{table}[h!]
\caption{Contingency table for participants in ELAs and national population, by gender. Pearson's Chi-squared test results for gender : X-squared = 277.61,  p-value $<$ 2.2$\cdot 10^{-6}$ }
\label{tab:sex_chi}
\centering
\begin{tabular}{|l|c|c|}
\hline 
 & Female & Male \\ 
\hline 
National population & 7361978 & 6924092 \\ 
Participants in ELAs & 56393 & 47803 \\ 
\hline 
\end{tabular}
\end{table}

\pagebreak
\clearpage

\subsection{Geographical representativeness}

\begin{figure}[!h]
\centering
\includegraphics[width=\textwidth]{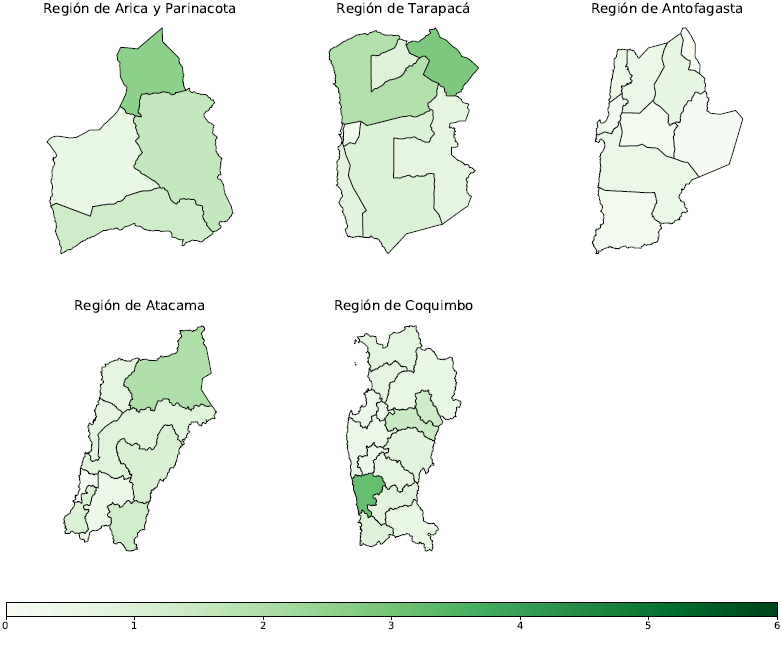}
\caption{Percentage of population participating in ELAs, by municipality (North).}
\label{fig:norte}
\end{figure}

\begin{figure}[!h]
\centering
\includegraphics[width=\textwidth]{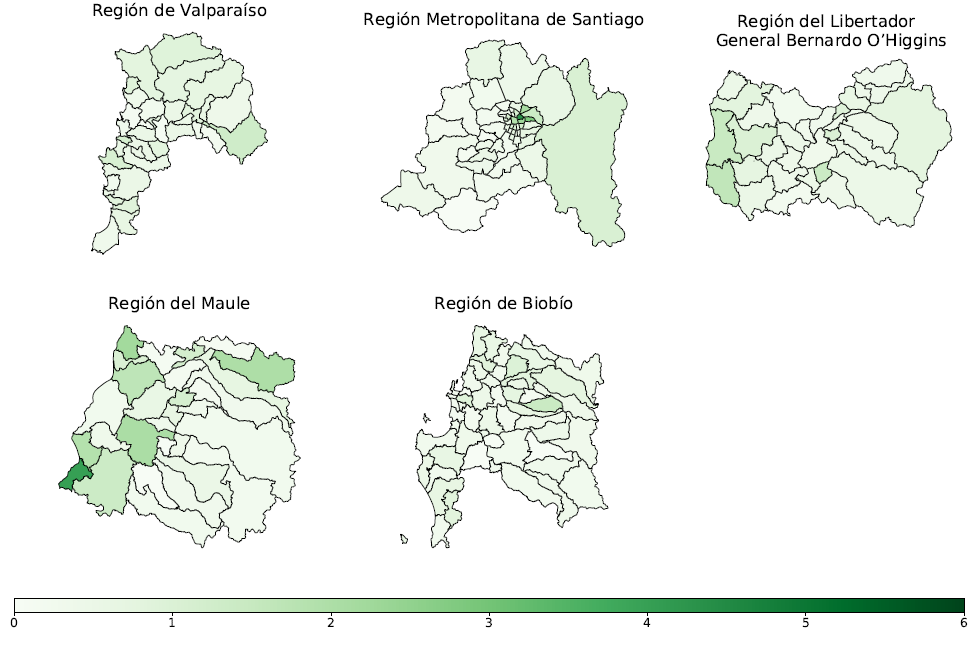}
\caption{Percentage of population participating in ELAs, by municipality (Central Chile).}
\label{fig:centro}
\end{figure}

\begin{figure}[!h]
\centering
\includegraphics[width=\textwidth]{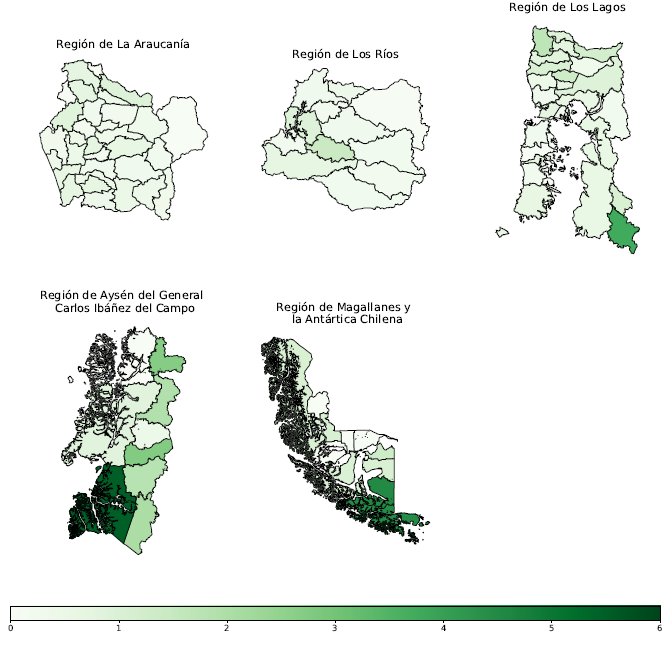}
\caption{Percentage of population participating in ELAs, by municipality (South).}
\label{fig:sur}
\end{figure}

\pagebreak
\clearpage

\subsection{ELAs : Questions and concepts.}\label{sec:elasqq}
What should be the fundamental and universal RIGHTS contained in the Constitution? Choose up to seven topics among the list below or suggest others in the free space. 

\begin{table}[!htbp]
\centering
\caption{ELAs : Questions and concepts.} 
 \label{tab:qqcc} 
\begin{tabular}{ll}
\hline
\hline \\[-1.8ex] 
\textit{Original concepts:} & \\
\hline \\[-1.8ex] 
Suffrage/vote               &    Honor/reputation  \\
Nationality                  &    Right of association \\
Election to public office    &   Peaceful assembly \\
Participation                & Request before the authorities \\
Life                        & Freedom to work \\
Mental and physical integrity    & Freedom of Education \\
Security/non-violence       &   Right to Work \\
Equality                    &   Fair wage \\
Non-discrimination          & Decent housing \\
Equality before the law     & Healthcare \\
Access to justice/due process   & Education \\
Equality in relation to public burdens   & Social security \\
Tax equality            &    Right to organize and to collective bargaining \\
Gender equity           & Right to strike \\
Children's and teenager’s rights  & Access to culture \\
Integration of disabled people    & Cultural identity \\
Personal freedom              & Indigenous people \\
Freedom of movement          & Environmental respect/protection \\
Freedom of conscience      &   Property \\
Freedom of expression        & Judicial protection of individual rights \\
Right to information       & Free economic initiative/free enterprise \\
Access to public information     &  None\\  
Privacy and intimacy      &  Others, specify\\
\hline \\[-1.8ex] 
\textit{New concepts:} & \\
\hline \\[-1.8ex] 
Standard of living  &  Cultural identity of indigenous people \\
 & Freedom of worship \\
Respect life from conception & Right to water \\
Right to make one's own decisions about one's life & Freedom \\
Right to work and a decent wage & Human Rights \\
Social Rights & Freedom of information and speech \\
Animal rights & Conservation of cultural and historical heritage\\
\hline
\hline
\end{tabular}
\end{table}

\pagebreak
\clearpage

\subsection{Statistical model: Size of ELAs}

\begin{figure}[!h]
\centering
\includegraphics[width=0.9\linewidth]{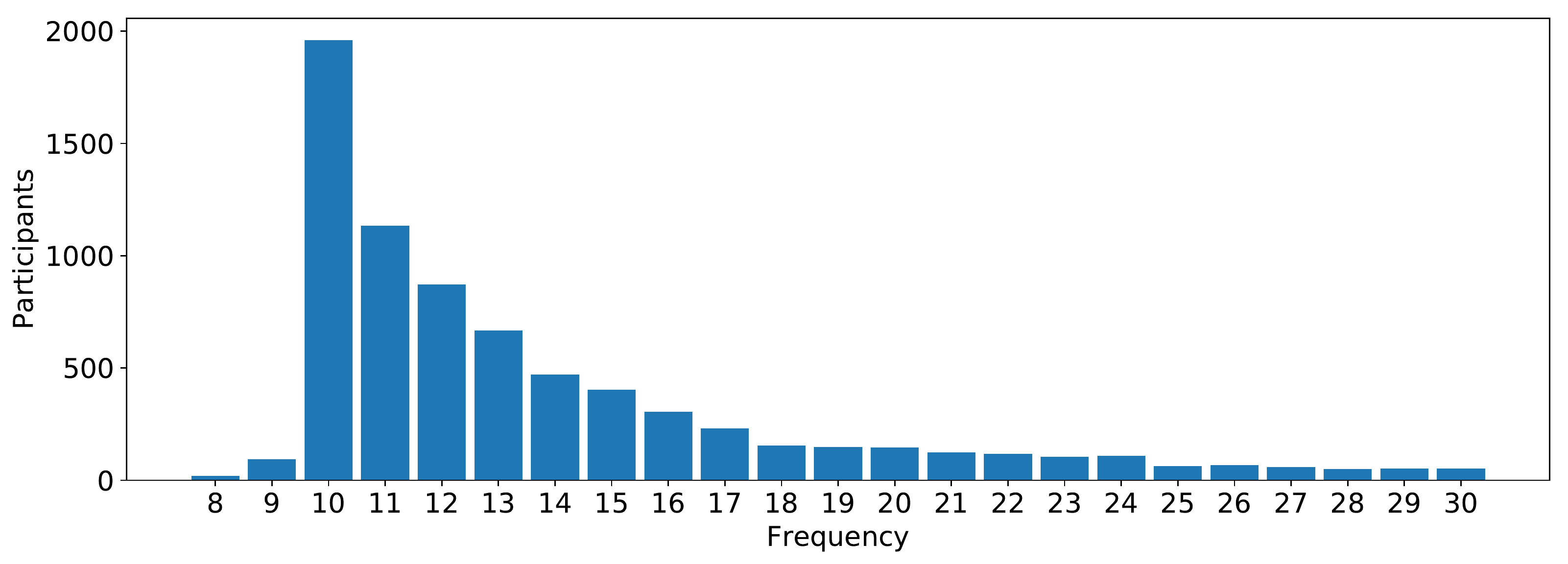}
\caption{Histogram of ELAs number of participants.}
\label{fig:size_elas}
\end{figure}
  
\pagebreak
\clearpage

\subsection{Independent variables}\label{SMet2}

\begin{itemize}
\item Demographic variables: Population over 14 years old; population density; share of population living in rural areas; cohorts of age and gender; share of single-parent and two-parent families. 
\item Socioeconomic variables: Education, measured as the share of people with higher education; internet penetration rate; socio-demographic classification in 7 groups, where group 1 is the most vulnerable in terms of its dependence on the municipal common fund and the local population. This classification was developed by SUBDERE in 2005, and it is incorporated in the model as a categorical variable. We also use a measure of poverty, obtained with a Small Area Estimation\footnote{Small-area estimation (SAE) refers to methods to address the limitations of survey data to produce reliable estimates of poverty for different geographical locations. The methodology used in Chile combines data from CASEN curvey and Census.} (SAE) methodology, as an alternative for SUBDERE groups. In Section~\ref{sec:ccs} we use a socioeconomic development index (SEDI) proposed by the Public Health Observatory in Chile, which comprises education, income, poverty and housing materiality. 
\item Social capital variables: number of community organizations in the municipality, such as parent centers, cultural centers, sport clubs, among others; share of municipality population that declares to participate in a community organization.
\item Political variables: Party affiliation, measured as the share of population affiliated to a political party; voter turnout in 2013's presidential elections; share of votes received by the winning candidate in 2013's presidential elections, who started the constitutional process in 2015 and was the standing president in 2016; share of population employed by the municipality (municipal officials). Additionally, three dummy variables were include to measure the effect of the  political party which supported the winning candidate for mayor, during the 2012's  municipal elections: the first one is equal to 1 if the party is within the government coalition, and 0 otherwise; the second one is equal to 1 if the party is in the opposition, and 0 otherwise. The third dummy variable is assigned to 1 when the mayor ran for office with no formal party support. Finally, a dummy variable is included for incumbent mayors. 
\item Religion: In Chile, evidence suggest that participation in evangelical Christians religion can provide skills that members can transfer to political activity \cite{patterson2005religious}. Therefore, an additional variable is added to account for the share of evangelical Christians living in the municipality. 

\item Internet penetration rate: This variable has been derived from CASEN survey. To assess the real impact of CASEN's lack of representativeness on the Internet effect, we performed the following comparison: (i) full model with all the 328 counties comprised in CASEN survey; (ii) full model with the 139 counties where the CASEN survey is representative; (iii) bootstrapping of the full model with random samples of 139 counties (the actual number of observations in each case is slightly different due to missing values). The purpose of (iii) is to evaluate whether the changes from (i) to (ii) result from the reduction of the sample size. The result of this comparison is shown in Table~\ref{tab:SM_casen}.
\end{itemize}


\begin{table}[!htbp] \centering 
  \caption{Descriptive statistics} 
  \label{tab:stats} 
\begin{tabular}{@{\extracolsep{5pt}}lccccc} 
\\[-1.8ex]\hline 
\hline \\[-1.8ex] 
Statistic & \multicolumn{1}{c}{N} & \multicolumn{1}{c}{Mean} & \multicolumn{1}{c}{St. Dev.} & \multicolumn{1}{c}{Min} & \multicolumn{1}{c}{Max} \\ 
\hline \\[-1.8ex] 
Number of ELAs  & 345 & 22.31 & 52.65 & 0 & 530 \\ 
Number of participants & 345 & 302.02 & 705.62 & 0 & 7,233 \\ 
Population & 345 & 41,408.90 & 64,527.83 & 230 & 453,530 \\
Population density  & 345 & 953.44 & 2,989.08 & 0.03 & 17,144.86 \\ 
Rurality & 345 & 0.36 & 0.29 & 0.00 & 1.00 \\ 
Higher education & 345 & 0.19 & 0.10 & 0.06 & 0.76 \\ 
Number of organizations & 342 & 472.99 & 599.02 & 3.00 & 4,892.00 \\ 
Participation in comm. org. & 324 & 0.07 & 0.08 & 0.0000 & 0.65 \\ 
Internet penetration rate & 324 & 0.05 & 0.12 & $-$0.20 & 0.90 \\ 
SEDI & 324 & 0.54 & 0.12 & 0.24 & 0.99 \\ 
Poverty & 344 & 0.13 & 0.08 & 0.001 & 0.42 \\
Born in 1981 or after & 345 & 0.40 & 0.05 & 0.29 & 0.60 \\ 
Men & 345 & 0.51 & 0.06 & 0.44 & 0.86 \\ 
Women & 345 & 0.49 & 0.06 & 0.14 & 0.56 \\ 
Single-parent family with children & 345 & 0.12 & 0.02 & 0.01 & 0.17 \\ 
Two-parent family with children & 345 & 0.28 & 0.05 & 0.07 & 0.45 \\ 
Voter turnout  & 345 & 0.50 & 0.08 & 0.15 & 0.68 \\ 
Votes for current president & 345 & 0.52 & 0.09 & 0.13 & 0.73 \\ 
Voter turnout (runoff) & 345 & 0.43 & 0.08 & 0.10 & 0.63 \\
Votes for current president (runoff) & 345 & 0.64 & 0.09 & 0.18 & 0.82 \\ 
Party affiliation & 345 & 0.08 & 0.07 & 0.01 & 0.94 \\ 
Municipal officials & 345 & 0.01 & 0.01 & 0.001 & 0.09 \\ 
Evangelical Christians & 341 & 0.13 & 0.08 & 0.01 & 0.49 \\ 
\hline \\[-1.8ex] 
\end{tabular} 
\end{table} 

\pagebreak
\clearpage

\pagebreak
\clearpage

\subsection{Text analyzing}\label{SMet3}

The analysis of large text corpora has proven fruitful in a range of topics in social sciences, including the analysis of discourse surrounding social movements \cite{bail2012}, content analysis for political texts \cite{grimmer2010,grimmer2013}, mapping social conflicts from legal sentences \cite{herrera2019understanding}, research choice and the production of scientific knowledge \cite{foster2015}, and communication and collaboration within organizations\cite{goldberg2016} (for a review of the history of content and text analysis in sociology and the social sciences, see Evans and Aceves (2016) \cite{evans2016}). In particular, Topic Modeling refers to a group of inductive techniques used to discover hidden topics contained in text documents. In general, these models assume that each document in a text corpus is produced from a mixture of latent topics, which in turn consists of a collection of words. Among these models, Structural Topic Modeling (STM) is a promising technique that enables researchers to incorporate document-level metadata information. STM has been used to find latent topics in social scientific research \cite{lindstedt2019, bohr2018}, to analyze open-ended survey responses \cite{roberts2014}, and even to uncover the underlying network structure of the radical activist groups by identifying topics in radical environmentalist publications \cite{almquist2019}. To implement this analysis we use the standard \textit{stm} R package developed by Roberts, Stewart, and Tingley (2018) \cite{stm}.

It is useful to define certain terminology specific to topic modeling \cite{Blei2012}.
      
\begin{itemize}
\item Topics: variables not observed in our data set; latent variables that allow capturing abstract notion; defined as the set of elements that can represent a theme present in a collection of documents without loss of statistical information.
\item Word: is the basic unit of discrete data, defined to be an element of a vocabulary indexed by $ {1 \dots V} $. Words are represented by unit-based vectors that have a single component equal to 1 and all others equal to zero. By means of supra-indexes the vocabulary components are denoted, for example, the $ v ^{th}$ indexed word is represented by a vector $ w $ long $ V $ such that $ w^v = 1 $ and $ w^u = 0 $ for $ u \neq v $.
\item Document: is a sequence of $ N $ words denoted by $ w = (w_1, w_2 \dots w_n) $, where $ w_n $ is the $ n-th $ word in the sequence. In our particular context, a document corresponds to a online post.

\item Corpus: is a collection of $ M $ documents denoted by $ D = {w_1, w_2, \dots w_M} $.
\end{itemize}

\paragraph*{CTM}  

Correlation Topic Model (CTM) is generative probabilistic model of a corpus. The basic idea is that documents are represented as random mixtures over latent topics, whereby each topic is characterized by a distribution over words \cite{blei2003}. CTM uses a logistic normal distribution for the topic proportions, which allows for covariance structure among the components. This distribution models correlation between components of the simplicial random variable through the covariance matrix of the normal distribution. 

\vspace{2mm}

Let $\{ \mu, \Sigma \}$ be a $K$-dimensional mean and covariance matrix, and let topics $\beta_{1:K}$ be $K$ multinomial distributions over a fixed word vocabulary. The generative process for an $N$-word documents is as follows:\cite{blei2006ctm}

\begin{enumerate}
    \item Draw topic proportions $\eta | \{ \mu, \Sigma \} \sim  \text{LogisticNormal} (\mu, \Sigma)$.
    \item For each word $n$ in \{1...N\}
    \begin{itemize}
        \item Draw topic assignment $z_n | \eta $ from Mult($f(\eta)$) 
        \item Draw word from that topic, $w_n | z_n,\beta_{1:K} $ from Mult($\beta_{z_n}$)
    \end{itemize}
\end{enumerate}

where $f(\eta_i) = exp(\eta_i)/\sum_j exp(\eta_j)$.

\paragraph*{STM}

Structural Topic Model (STM) is a general framework for topic modeling, which enables users to incorporate document-level covariate information. The goal of STM is to discover topics and estimate their relationship to document metadata, conducting hypothesis testing about these relationships. Thus, this method serves as bridge between statistical techniques and research goals. Like CTM, STM is a generative model of word counting, where a topic is defined as a mixture over words where each word has a probability of belonging to a topic. The \textit{stm} R package implements the estimation algorithms for the model \cite{stm}.  In the case of no covariates, the model reduces to a fast implementation of CTM. The document metadata can be incorporated in two ways: as topical prevalence, which refers to how much of a document is associated with a topic, and as topical content, related to the words used within a topic. The generative process is similar to CTM but incorporating the $p$ covariates in a 1-by-$p$ vector of document covariates $X_d$ for document $d$. 

\begin{enumerate}
    \item Draw topic distribution $\theta_d | X_d \gamma, \Sigma  \sim$ LogisticNormal($\mu = X_d \gamma, \Sigma$), where $\gamma$ is a $p$-by-$K-1$ matrix of coefficients and $\Sigma$ is $K-1$-by-$K-1$ covariance matrix.
    \item The probability on, per topic, word distribution is $\beta_{d,k}$. Given a document-level content variable $y_d$, the document-specific distribution $\beta_{d,k}$ is formed over words representing each topic ($k$) using the baseline word distribution ($m$), the topic specific deviation $\kappa_k$, the covariate group deviation $\kappa_{y_d}$ and the interaction between the two $\kappa_{y_d,k}$
    \begin{equation}
        \beta_{d,k} \propto e^{m + \kappa_k + \kappa_{y_d} + \kappa_{y_d,k} }
    \end{equation}
    \item  For each word $n$ in \{1...N\}
    \begin{itemize}
        \item Draw topic assignment $z_n | \theta_d $ from Mult($\theta_d$) 
        \item Draw word from that topic, $w_n |  z_n,\beta_{1:K} $ from Mult($\beta_{z_n}$)
    \end{itemize}
\end{enumerate}

\clearpage
\newpage

\section{Supplementary results}
\label{sec:SMresults}

\begin{table}[!htbp] 
\centering 
  \caption{Variance inflation factor (VIF) for predictors in OLS model.} 
  \label{tab:SM_VIF} 
\begin{tabular}{@{\extracolsep{5pt}}lc } 
\\[-1.8ex]\hline 
\hline \\[-1.8ex] 
Regressor &          VIF  \\
                     \hline \\[-1.8ex] 
Log (population)    & 10.3908 \\         
Higher education      &     5.5627         \\
Internet penetration rate  &              2.0860    \\     
Log (community organizations) &           3.0870   \\
Born in 1981 or after          &     3.0040\\
Rurality    &     3.0160         \\
Log (population density) &      3.4063\\         
Women   &        2.5619         \\
Two-parent family (with children) &  1.9531         \\
Single-parent family (with children) &        1.4367  \\       
Votes for current president   & 2.9716         \\
Municipal officials      & 3.1048         \\
Voter turnout        & 2.4574         \\
Party affiliation  & 1.7249         \\
Evangelical Christians &             1.2018\\         
\hline 
\hline \\[-1.8ex] 
\end{tabular} 
\end{table}  


\begin{figure}[h!]
\centering
\includegraphics[width=5in]{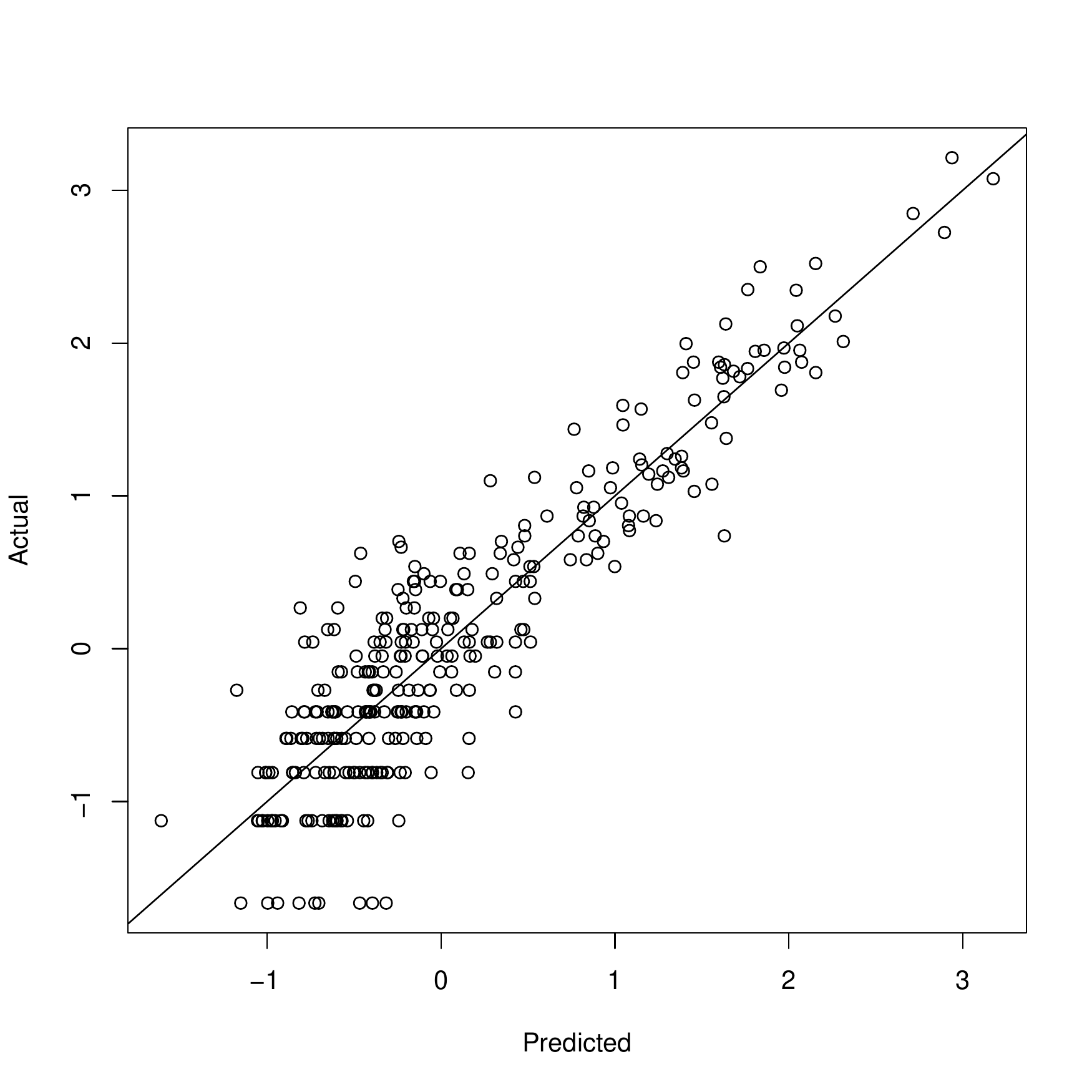}
\caption{Actual versus Predicted values for the number of ELAs, model (3) of Table~\ref{tab:all}.}
\label{fig_sim}
\end{figure}

\clearpage
\newpage

\begin{figure}[h!]
\centering
\includegraphics[width=0.9\linewidth]{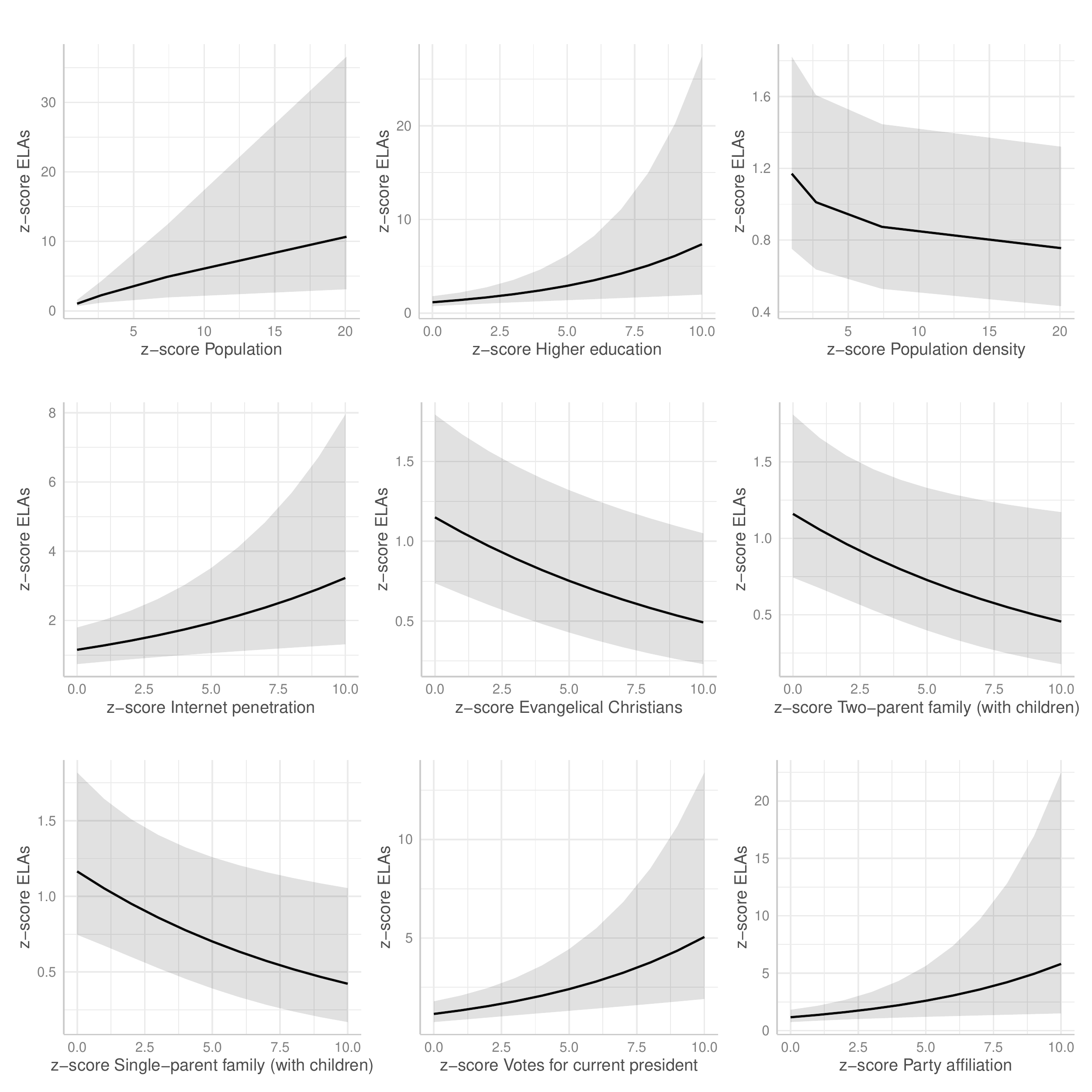}
\caption{Marginal effects of model predictors, for model (3) of Table~\ref{tab:all}. Bands reflect $95\%$ confidence intervals.}
\label{fig_marginals}
\end{figure}

\clearpage
\newpage

\begin{table}[!htbp] 
\scriptsize
\centering 
  \caption{OLS estimates for the  standardized full model with robust standard errors, p-value RESET test = 0.3501. RESET test were performed on the second power of regressors.} 
  \label{tab:SM_full} 
\begin{tabular}{@{\extracolsep{5pt}}lD{.}{.}{-3} } 
\\[-1.8ex]\hline 
\hline \\[-1.8ex] 
 & \multicolumn{1}{c}{\textit{Dependent variable:}} \\ 
\cline{2-2} 
\\[-1.8ex] & \multicolumn{1}{c}{log (1 + ELAs)} \\ 
\hline \\[-1.8ex] 
  Log (population) & 0.778^{***} \\ 
  & (0.164) \\ 
  Higher education & 0.185^{***} \\ 
  & (0.065) \\ 
  Internet penetration rate & 0.103^{***} \\ 
  & (0.039) \\ 
    \textit{SUBDERE groups} &\\
  & \\[-1.8ex] 
  \quad Group 2 & -0.122 \\ 
  & (0.154) \\ 
  \quad Group 3 & -0.234 \\ 
  & (0.196) \\ 
  \quad Group 4 & -0.129 \\ 
  & (0.224) \\ 
  \quad Group 5 & 0.016 \\ 
  & (0.277) \\ 
  \quad Group 6 & 0.297 \\ 
  & (0.351) \\ 
  \quad Group 7 & 0.497 \\ 
  & (0.417) \\ 
  \quad Group 8 & 0.236 \\ 
  & (0.421) \\ 
  Log (community organizations) & 0.085^{*} \\ 
  & (0.051) \\ 
  Born in 1981 or after & -0.008 \\ 
  & (0.056) \\ 
  Rurality & -0.052 \\ 
  & (0.054) \\ 
  Log (population density) & -0.139^{***} \\ 
  & (0.053) \\ 
  Women & 0.066 \\ 
  & (0.069) \\ 
  Two-parent family (with children) & -0.093^{**} \\ 
  & (0.043) \\ 
  Single-parent family (with children) & -0.101^{**} \\ 
  & (0.043) \\ 
  Votes for current president & 0.148^{***} \\ 
  & (0.046) \\ 
  Municipal officials & -0.040 \\ 
  & (0.170) \\ 
  Voter turnout & 0.087 \\ 
  & (0.054) \\ 
  Mayor (government) & 0.063 \\ 
  & (0.069) \\ 
  Mayor (opposition) & -0.099 \\ 
  & (0.079) \\ 
  Party affiliation & 0.161^{**} \\ 
  & (0.066) \\ 
  Incumbent mayor (True) & -0.020 \\ 
  & (0.054) \\ 
  Evangelical Christians & -0.075^{***} \\ 
  & (0.027) \\ 
  \hline\\[-1.8ex]
  Rurality * Evangelical Christians & 0.040 \\ 
  & (0.049) \\ 
  Log (population density) * Evangelical Christians & -0.055 \\ 
  & (0.067) \\ 
  Log (community organizations) * Born in 1981 or after & -0.054 \\ 
  & (0.043) \\ 
  Born in 1981 or after * Internet penetration rate & 0.005 \\ 
  & (0.018) \\ 
  Rurality * Log (community organizations)  & -0.053 \\ 
  & (0.050) \\ 
  Log (population density) * Log (community organizations) & -0.054 \\ 
  & (0.041) \\ 
  Municipal officials * Voter turnout & 0.039 \\ 
  & (0.095) \\ 
  Municipal officials * Mayor (government) & 0.204 \\ 
  & (0.181) \\ 
  Municipal officials * Mayor (opposition) & -0.027 \\ 
  & (0.184) \\ 
  Constant & 0.037 \\ 
  & (0.209) \\ 
 \hline \\[-1.8ex] 
Observations & \multicolumn{1}{c}{310} \\ 
Adjusted R$^{2}$ & \multicolumn{1}{c}{0.814} \\ 
Residual Std. Error & \multicolumn{1}{c}{0.424 (df = 275)} \\ 
F Statistic & \multicolumn{1}{c}{40.652$^{***}$ (df = 34; 275)} \\ 
\hline 
\hline \\[-1.8ex] 
\multicolumn{2}{p{10cm}}{Note: $^{*}$p$<$0.1; $^{**}$p$<$0.05; $^{***}$p$<$0.01. The base categories for dummy variables are:  ``Group 1'' for SUBDERE groups, ``False'' for Incumbent Mayor an ``Independent'' for Mayor. See Appendix Table Al for more detailed variable definitions and sources}
\end{tabular} 
\end{table} 

\clearpage
\newpage

\begin{table}[!htbp] \centering 
\scriptsize
  \caption{OLS and Negative Binomial estimates for the full model.} 
  \label{tab:SM_poisson} 
\begin{tabular}{@{\extracolsep{5pt}}lD{.}{.}{-3} D{.}{.}{-3} } 
\\[-1.8ex]\hline 
\hline \\[-1.8ex] 
 & \multicolumn{2}{c}{\textit{Dependent variable:}} \\ 
\cline{2-3} 
\\[-1.8ex] & \multicolumn{1}{c}{OLS} & \multicolumn{1}{c}{Negative binomial} \\ 
\\[-1.8ex] 
\\[-1.8ex] & \multicolumn{1}{c}{log (1 + ELAs)} & \multicolumn{1}{c}{ELAs} \\ 
\hline \\[-1.8ex] 
  Log(population) & 0.778^{***} & 1.263^{***} \\ 
  & (0.164) & (0.218) \\ 
  Higher education & 0.185^{***} & 0.282^{***} \\ 
  & (0.065) & (0.099) \\ 
  Internet penetration rate & 0.103^{***} & 0.116^{**} \\ 
  & (0.039) & (0.054) \\ 
    \textit{SUBDERE groups} &\\
  & \\[-1.8ex]  
  \quad  Group 2 & -0.122 & -0.221 \\ 
  & (0.154) & (0.285) \\ 
  \quad Group 3 & -0.234 & -0.403 \\ 
  & (0.196) & (0.336) \\ 
  \quad Group 4 & -0.129 & -0.327 \\ 
  & (0.224) & (0.368) \\ 
  \quad Group 5 & 0.016 & -0.212 \\ 
  & (0.277) & (0.421) \\ 
  \quad Group 6 & 0.297 & 0.038 \\ 
  & (0.351) & (0.507) \\ 
  \quad Group 7 & 0.497 & 0.222 \\ 
  & (0.417) & (0.570) \\ 
  \quad Group 8 & 0.236 & -0.142 \\ 
  & (0.421) & (0.572) \\ 
  Log (community organizations)  & 0.085^{*} & 0.128^{*} \\ 
  & (0.051) & (0.077) \\ 
  Born in 1981 or after & -0.008 & -0.072 \\ 
  & (0.056) & (0.085) \\ 
  Rurality & -0.052 & -0.119 \\ 
  & (0.054) & (0.083) \\ 
  Log (population density) & -0.139^{***} & -0.238^{***} \\ 
  & (0.053) & (0.073) \\ 
  Women & 0.066 & 0.045 \\ 
  & (0.069) & (0.098) \\ 
  Two-parent family (with children) & -0.093^{**} & -0.135^{**} \\ 
  & (0.043) & (0.059) \\ 
  Single-parent family (with children) & -0.101^{**} & -0.146^{**} \\ 
  & (0.043) & (0.065) \\ 
  Votes for current president & 0.148^{***} & 0.231^{***} \\ 
  & (0.046) & (0.065) \\ 
  Municipal officials & -0.040 & 0.008 \\ 
  & (0.170) & (0.279) \\ 
  Voter turnout & 0.087 & 0.135 \\ 
  & (0.054) & (0.090) \\ 
  Mayor (government) & 0.063 & 0.072 \\ 
  & (0.069) & (0.108) \\ 
  Mayor (opposition) & -0.099 & -0.211 \\ 
  & (0.079) & (0.129) \\ 
  Party affiliation & 0.161^{**} & 0.239^{**} \\ 
  & (0.066) & (0.096) \\ 
  Incumbent mayor (True) & -0.020 & -0.004 \\ 
  & (0.054) & (0.073) \\ 
  Evangelical Christians & -0.075^{***} & -0.084^{**} \\ 
  & (0.027) & (0.041) \\ 
    \hline\\[-1.8ex]
  Log(population density) * Evangelical Christians & -0.055 & -0.134^{*} \\ 
  & (0.067) & (0.081) \\ 
  Constant & 0.037 & 2.308^{***} \\ 
  & (0.209) & (0.340) \\ 
 \hline \\[-1.8ex] 
Observations & \multicolumn{1}{c}{310} & \multicolumn{1}{c}{310} \\ 
Adjusted R$^{2}$ & \multicolumn{1}{c}{0.814} &  \\ 
Log Likelihood &  & \multicolumn{1}{c}{-930.405} \\ 
$\theta$ &  & \multicolumn{1}{c}{$6.338^{***}$  (0.893)} \\ 
Akaike Inf. Crit. &  & \multicolumn{1}{c}{1,930.810} \\ 
Residual Std. Error & \multicolumn{1}{c}{0.424 (df = 275)} &  \\ 
F Statistic & \multicolumn{1}{c}{40.652$^{***}$ (df = 34; 275)} &  \\ 
\hline 
\hline \\[-1.8ex] 
\multicolumn{3}{p{10cm}}{Note: $^{*}$p$<$0.1; $^{**}$p$<$0.05; $^{***}$p$<$0.01. The base categories for dummy variables are:  ``Group 1'' for SUBDERE groups, ``False'' for Incumbent Mayor an ``Independent'' for Mayor. Only significant interactions are shown. For both models, all independent variables have been standardized. The dependent variable (\textit{ELAs}) is also standardize in the OLS model, whereas it is kept as a count variable in the Negative Binomial model.}
\end{tabular} 
\end{table}

\newpage

\begin{table}[!htbp] \centering 
\scriptsize
  \caption{OLS estimates for the number of ELAs (log (1 + ELAs), p-value RESET test = 0.3501) and the number of participants (log (1 + participants), p-value RESET test = 0.4672). RESET tests were performed on the second power of regressors.} 
  \label{tab:ols_pob} 
\begin{tabular}{@{\extracolsep{5pt}}lD{.}{.}{-3} D{.}{.}{-3} } 
\\[-1.8ex]\hline 
\hline \\[-1.8ex] 
 & \multicolumn{2}{c}{\textit{Outcome variable:}} \\ 
\cline{2-3} 
\\[-1.8ex] & \multicolumn{1}{c}{log (1 + ELAs)} & \multicolumn{1}{c}{log (1 + participants)} \\ 
\hline \\[-1.8ex] 
 Log(population) & 0.778^{***} & 0.747^{***} \\ 
  & (0.164) & (0.181) \\ 
  Higher education & 0.185^{***} & 0.148^{**} \\ 
  & (0.065) & (0.073) \\ 
  Internet penetration rate & 0.103^{***} & 0.114^{**} \\ 
  & (0.039) & (0.045) \\ 
  \textit{SUBDERE groups} &&\\
  && \\[-1.8ex] 
  \quad Group 2 & -0.122 & -0.067 \\ 
  & (0.154) & (0.210) \\ 
  \quad Group 3 & -0.234 & -0.153 \\ 
  & (0.196) & (0.261) \\ 
  \quad Group 4 & -0.129 & 0.020 \\ 
  & (0.224) & (0.290) \\ 
  \quad Group 5 & 0.016 & 0.154 \\ 
  & (0.277) & (0.352) \\ 
  \quad Group 6 & 0.297 & 0.250 \\ 
  & (0.351) & (0.439) \\ 
  \quad Group 7 & 0.497 & 0.385 \\ 
  & (0.417) & (0.500) \\ 
  \quad Group 8 & 0.236 & 0.044 \\ 
  & (0.421) & (0.501) \\ 
  Log (community organizations) & 0.085^{*} & 0.103^{*} \\ 
  & (0.051) & (0.061) \\ 
  Born in 1981 or after & -0.008 & 0.041 \\ 
  & (0.056) & (0.066) \\ 
  Rurality & -0.052 & 0.014 \\ 
  & (0.054) & (0.061) \\ 
  Log(population density) & -0.139^{***} & -0.076 \\ 
  & (0.053) & (0.063) \\ 
  Women & 0.066 & 0.017 \\ 
  & (0.069) & (0.095) \\ 
  Two-parent family (with children) & -0.093^{**} & -0.100^{**} \\ 
  & (0.043) & (0.050) \\ 
  Single-parent family (with children) & -0.101^{**} & -0.078 \\ 
  & (0.043) & (0.053) \\ 
  Votes for current president & 0.148^{***} & 0.135^{**} \\ 
  & (0.046) & (0.056) \\ 
  Municipal officials & -0.040 & 0.024 \\ 
  & (0.170) & (0.194) \\ 
  Voter turnout & 0.087 & 0.178^{**} \\ 
  & (0.054) & (0.075) \\ 
  Mayor (government) & 0.063 & 0.095 \\ 
  & (0.069) & (0.088) \\ 
  Mayor (opposition) & -0.099 & -0.136 \\ 
  & (0.079) & (0.100) \\ 
  Party affiliation & 0.161^{**} & 0.200^{**} \\ 
  & (0.066) & (0.078) \\ 
  Incumbent mayor (True) & -0.020 & -0.005 \\ 
  & (0.054) & (0.067) \\ 
  Evangelical Christians & -0.075^{***} & -0.071^{**} \\ 
  & (0.027) & (0.033) \\ 
  \hline\\[-1.8ex] 
  Constant & 0.037 & 0.008 \\ 
  & (0.209) & (0.272) \\ 
 \hline \\[-1.8ex] 
Observations & \multicolumn{1}{c}{310} & \multicolumn{1}{c}{310} \\ 
Adjusted R$^{2}$ & \multicolumn{1}{c}{0.814} & \multicolumn{1}{c}{0.698} \\ 
Residual Std. Error (df = 275) & \multicolumn{1}{c}{0.424} & \multicolumn{1}{c}{0.511} \\ 
F Statistic (df = 34; 275) & \multicolumn{1}{c}{40.652$^{***}$} & \multicolumn{1}{c}{22.046$^{***}$} \\ 
\hline 
\hline \\[-1.8ex] 
\multicolumn{3}{p{10cm}}{Note: $^{*}$p$<$0.1; $^{**}$p$<$0.05; $^{***}$p$<$0.01. The base categories for dummy variables are:  ``Group 1'' for SUBDERE groups, ``False'' for Incumbent Mayor an ``Independent'' for Mayor. Only significant interactions are shown.}
\end{tabular} 
\end{table}

\begin{table}[!htbp] \centering 
\scriptsize
  \caption{OLS estimates for the full model (p-value RESET test = 0.3501), and replacing \textit{SUBDERE groups} with \textit{poverty} (p-value RESET test = 0.2828). RESET tests were performed on the second power of regressors.} 
  \label{tab:SM_pobr} 
\begin{tabular}{@{\extracolsep{5pt}}lD{.}{.}{-3} D{.}{.}{-3} } 
\\[-1.8ex]\hline 
\hline \\[-1.8ex] 
 & \multicolumn{2}{c}{\textit{Outcome variable:}} \\ 
\cline{2-3} 
\\[-1.8ex] & \multicolumn{2}{c}{log (1 + ELAs)} \\ 
\\[-1.8ex] & \multicolumn{1}{c}{SUBDERE Groups} & \multicolumn{1}{c}{Poverty}\\ 
\hline \\[-1.8ex] 
  Log(population) & 0.778^{***} & 0.971^{***} \\ 
  & (0.164) & (0.112) \\ 
  Higher education & 0.185^{***} & 0.248^{***} \\ 
  & (0.065) & (0.071) \\ 
  Internet penetration rate & 0.103^{***} & 0.071^{*} \\ 
  & (0.039) & (0.043) \\ 
  \textit{SUBDERE groups} &&\\
  && \\[-1.8ex] 
  \quad Group 2 & -0.122 &  \\ 
  & (0.154) &  \\ 
  \quad Group 3 & -0.234 &  \\ 
  & (0.196) &  \\ 
  \quad Group 4 & -0.129 &  \\ 
  & (0.224) &  \\ 
  \quad Group 5 & 0.016 &  \\ 
  & (0.277) &  \\ 
  \quad Group 6 & 0.297 &  \\ 
  & (0.351) &  \\ 
  \quad Group 7 & 0.497 &  \\ 
  & (0.417) &  \\ 
  \quad Group 8 & 0.236 &  \\ 
  & (0.421) &  \\ 
  Poverty &  & 0.033 \\ 
  &  & (0.041) \\ 
  Log (community organizations) & 0.085^{*} & 0.085^{*} \\ 
  & (0.051) & (0.051) \\ 
  Born in 1981 or after & -0.008 & 0.003 \\ 
  & (0.056) & (0.047) \\ 
  Rurality & -0.052 & -0.028 \\ 
  & (0.054) & (0.056) \\ 
  Log (population density) & -0.139^{***} & -0.124^{**} \\ 
  & (0.053) & (0.052) \\ 
  Women & 0.066 & 0.077 \\ 
  & (0.069) & (0.073) \\ 
  Two-parent family (with children) & -0.093^{**} & -0.106^{**} \\ 
  & (0.043) & (0.042) \\ 
  Single-parent family (with children) & -0.101^{**} & -0.094^{**} \\ 
  & (0.043) & (0.048) \\ 
  Votes for current president & 0.148^{***} & 0.153^{***} \\ 
  & (0.046) & (0.045) \\ 
  Municipal officials & -0.040 & 0.013 \\ 
  & (0.170) & (0.146) \\ 
  Voter turnout & 0.087 & 0.065 \\ 
  & (0.054) & (0.055) \\ 
  Mayor (government) & 0.063 & 0.120^{*} \\ 
  & (0.069) & (0.068) \\ 
  Mayor (opposition) & -0.099 & -0.036 \\ 
  & (0.079) & (0.078) \\ 
  Party affiliation & 0.161^{**} & 0.217^{***} \\ 
  & (0.066) & (0.068) \\ 
  Incumbent mayor (True) & -0.020 & -0.013 \\ 
  & (0.054) & (0.052) \\ 
  Evangelical Christians & -0.075^{***} & -0.087^{***} \\ 
  & (0.027) & (0.031) \\ 
  \hline\\[-1.8ex] 
  Municipal officials * Mayor (government) & 0.204 & 0.295^{*} \\ 
  & (0.181) & (0.172) \\ 
  Constant & 0.037 & -0.043 \\ 
  & (0.209) & (0.055) \\ 
 \hline \\[-1.8ex] 
Observations & \multicolumn{1}{c}{310} & \multicolumn{1}{c}{320} \\ 
Adjusted R$^{2}$ & \multicolumn{1}{c}{0.814} & \multicolumn{1}{c}{0.810} \\ 
Residual Std. Error & \multicolumn{1}{c}{0.424 (df = 275)} & \multicolumn{1}{c}{0.432 (df = 291)} \\ 
F Statistic & \multicolumn{1}{c}{40.652$^{***}$ (df = 34; 275)} & \multicolumn{1}{c}{49.414$^{***}$ (df = 28; 291)} \\ 
\hline 
\hline \\[-1.8ex] 
\multicolumn{3}{p{10cm}}{Note: $^{*}$p$<$0.1; $^{**}$p$<$0.05; $^{***}$p$<$0.01. The base categories for dummy variables are:  ``Group 1'' for SUBDERE groups, ``False'' for Incumbent Mayor an ``Independent'' for Mayor. Only significant interactions are shown.}
\end{tabular} 
\end{table}

\begin{table}[!htbp] \centering 
\scriptsize
  \caption{OLS estimates for the full model (p-value RESET test = 0.3501) and replacing the votes for the incumbent president and the voter turnout with the corresponding runoff variables (p-value RESET test = 0.4605). RESET tests were performed on the second power of regressors.} 
  \label{tab:SM_votos} 
\begin{tabular}{@{\extracolsep{5pt}}lD{.}{.}{-3} D{.}{.}{-3} } 
\\[-1.8ex]\hline 
\hline \\[-1.8ex] 
 & \multicolumn{2}{c}{\textit{Dependent variable:}} \\ 
\cline{2-3} 
\\[-1.8ex] & \multicolumn{2}{c}{log (1 + ELAs)} \\ 
\\[-1.8ex] & \multicolumn{1}{c}{First round} & \multicolumn{1}{c}{Runoff}\\ 
\hline \\[-1.8ex] 
  Log(population) & 0.778^{***} & 0.763^{***} \\ 
  & (0.164) & (0.168) \\ 
  Higher education & 0.185^{***} & 0.131^{**} \\ 
  & (0.065) & (0.060) \\ 
  Internet penetration rate & 0.103^{***} & 0.111^{***} \\ 
  & (0.039) & (0.039) \\ 
  \textit{SUBDERE groups} &&\\
  && \\[-1.8ex]
  \quad Group 2 & -0.122 & -0.123 \\ 
  & (0.154) & (0.155) \\ 
  \quad Group 3 & -0.234 & -0.236 \\ 
  & (0.196) & (0.197) \\ 
  \quad Group 4 & -0.129 & -0.115 \\ 
  & (0.224) & (0.229) \\ 
  \quad Group 5 & 0.016 & 0.043 \\ 
  & (0.277) & (0.282) \\ 
  \quad Group 6 & 0.297 & 0.352 \\ 
  & (0.351) & (0.361) \\ 
  \quad Group 7 & 0.497 & 0.567 \\ 
  & (0.417) & (0.429) \\ 
  \quad Group 8 & 0.236 & 0.297 \\ 
  & (0.421) & (0.428) \\ 
  Log (community organizations) & 0.085^{*} & 0.085^{*} \\ 
  & (0.051) & (0.051) \\ 
  Born in 1981 or after & -0.008 & 0.006 \\ 
  & (0.056) & (0.061) \\ 
  Rurality & -0.052 & -0.048 \\ 
  & (0.054) & (0.055) \\ 
  Log (population density) & -0.139^{***} & -0.175^{***} \\ 
  & (0.053) & (0.053) \\ 
  Women & 0.066 & 0.065 \\ 
  & (0.069) & (0.070) \\ 
  Two-parent family (with children) & -0.093^{**} & -0.110^{**} \\ 
  & (0.043) & (0.046) \\ 
  Single-parent family (with children) & -0.101^{**} & -0.098^{**} \\ 
  & (0.043) & (0.043) \\ 
  Votes for current president & 0.148^{***} &  \\ 
  & (0.046) &  \\ 
  Votes for current president (runoff) &  & 0.089^{**} \\ 
  &  & (0.038) \\ 
  Municipal officials & -0.040 & -0.043 \\ 
  & (0.170) & (0.170) \\ 
  Voter turnout & 0.087 &  \\ 
  & (0.054) &  \\ 
  Voter turnout (runoff) &  & 0.155^{**} \\ 
  &  & (0.060) \\ 
  Mayor (government) & 0.063 & 0.064 \\ 
  & (0.069) & (0.069) \\ 
  Mayor (opposition) & -0.099 & -0.117 \\ 
  & (0.079) & (0.080) \\ 
  Party affiliation & 0.161^{**} & 0.161^{**} \\ 
  & (0.066) & (0.067) \\ 
  Incumbent mayor (True) & -0.020 & -0.022 \\ 
  & (0.054) & (0.054) \\ 
  Evangelical Christians & -0.075^{***} & -0.074^{***} \\ 
  & (0.027) & (0.028) \\ 
  \hline\\[-1.8ex]
  Constant & 0.037 & 0.016 \\ 
  & (0.209) & (0.213) \\ 
 \hline \\[-1.8ex] 
Observations & \multicolumn{1}{c}{310} & \multicolumn{1}{c}{310} \\ 
Adjusted R$^{2}$ & \multicolumn{1}{c}{0.814} & \multicolumn{1}{c}{0.813} \\ 
Residual Std. Error (df = 275) & \multicolumn{1}{c}{0.424} & \multicolumn{1}{c}{0.425} \\ 
F Statistic (df = 34; 275) & \multicolumn{1}{c}{40.652$^{***}$} & \multicolumn{1}{c}{40.479$^{***}$} \\ 
\hline 
\hline \\[-1.8ex] 
\multicolumn{3}{p{10cm}}{Note: $^{*}$p$<$0.1; $^{**}$p$<$0.05; $^{***}$p$<$0.01. The base categories for dummy variables are:  ``Group 1'' for SUBDERE groups, ``False'' for Incumbent Mayor an ``Independent'' for Mayor. Only significant interactions are shown.} 
\end{tabular}
\end{table}

\begin{table}[!htbp] \centering 
\scriptsize
  \caption{OLS estimates for the full model (p-value RESET test = 0.3501), and replacing the variables Mayor and Incumbent Mayor with a governments' influence variable (p-value RESET test = 0.6196). RESET tests were performed on the second power of regressors.} 
  \label{tab:SM_gov} 
\begin{tabular}{@{\extracolsep{5pt}}lD{.}{.}{-3} D{.}{.}{-3} } 
\\[-1.8ex]\hline 
\hline \\[-1.8ex] 
 & \multicolumn{2}{c}{\textit{Dependent variable:}} \\ 
\cline{2-3} 
\\[-1.8ex] & \multicolumn{2}{c}{log (1 + ELAs)} \\ 
\\[-1.8ex] & \multicolumn{1}{c}{Mayor} & \multicolumn{1}{c}{Gov. Influence}\\ 
\hline \\[-1.8ex] 
  Log(population) & 0.778^{***} & 0.775^{***} \\ 
  & (0.164) & (0.158) \\ 
  Higher education & 0.185^{***} & 0.195^{***} \\ 
  & (0.065) & (0.062) \\ 
  Internet penetration rate & 0.103^{***} & 0.097^{**} \\ 
  & (0.039) & (0.039) \\ 
    \textit{SUBDERE groups} &&\\
  && \\[-1.8ex]
  \quad Group 2 & -0.122 & -0.119 \\ 
  & (0.154) & (0.159) \\ 
  \quad Group 3 & -0.234 & -0.222 \\ 
  & (0.196) & (0.198) \\ 
  \quad Group 4 & -0.129 & -0.131 \\ 
  & (0.224) & (0.226) \\ 
  \quad Group 5 & 0.016 & 0.009 \\ 
  & (0.277) & (0.276) \\ 
  \quad Group 6 & 0.297 & 0.305 \\ 
  & (0.351) & (0.344) \\ 
  \quad Group 7 & 0.497 & 0.496 \\ 
  & (0.417) & (0.408) \\ 
  \quad Group 8 & 0.236 & 0.238 \\ 
  & (0.421) & (0.408) \\ 
  Log (community organizations) & 0.085^{*} & 0.083^{*} \\ 
  & (0.051) & (0.050) \\ 
  Born in 1981 or after & -0.008 & -0.012 \\ 
  & (0.056) & (0.056) \\ 
  Rurality & -0.052 & -0.065 \\ 
  & (0.054) & (0.053) \\ 
  Log (population density) & -0.139^{***} & -0.143^{***} \\ 
  & (0.053) & (0.053) \\ 
  Women & 0.066 & 0.055 \\ 
  & (0.069) & (0.071) \\ 
  Two-parent family (with children) & -0.093^{**} & -0.099^{**} \\ 
  & (0.043) & (0.044) \\ 
  Single-parent family (with children) & -0.101^{**} & -0.097^{**} \\ 
  & (0.043) & (0.043) \\ 
  Votes for current president & 0.148^{***} & 0.159^{***} \\ 
  & (0.046) & (0.045) \\ 
  Municipal officials & -0.040 & -0.005 \\ 
  & (0.170) & (0.149) \\ 
  Voter turnout & 0.087 & 0.077 \\ 
  & (0.054) & (0.053) \\ 
  Mayor (government) & 0.063 &  \\ 
  & (0.069) &  \\ 
  Mayor (opposition) & -0.099 &  \\ 
  & (0.079) &  \\ 
  Gov. influence &  & 0.027 \\ 
  &  & (0.032) \\ 
  Party affiliation & 0.161^{**} & 0.161^{**} \\ 
  & (0.066) & (0.068) \\ 
  Incumbent mayor (True) & -0.020 &  \\ 
  & (0.054) &  \\ 
  Evangelical Christians & -0.075^{***} & -0.069^{**} \\ 
  & (0.027) & (0.027) \\ 
  \hline\\[-1.8ex]
  Constant & 0.037 & 0.020 \\ 
  & (0.209) & (0.206) \\ 
 \hline \\[-1.8ex] 
Observations & \multicolumn{1}{c}{310} & \multicolumn{1}{c}{310} \\ 
R$^{2}$ & \multicolumn{1}{c}{0.834} & \multicolumn{1}{c}{0.831} \\ 
Adjusted R$^{2}$ & \multicolumn{1}{c}{0.814} & \multicolumn{1}{c}{0.813} \\ 
Residual Std. Error & \multicolumn{1}{c}{0.424 (df = 275)} & \multicolumn{1}{c}{0.425 (df = 278)} \\ 
F Statistic & \multicolumn{1}{c}{40.652$^{***}$ (df = 34; 275)} & \multicolumn{1}{c}{44.234$^{***}$ (df = 31; 278)} \\ 
\hline 
\hline \\[-1.8ex] 
\multicolumn{3}{p{10cm}}{Note: $^{*}$p$<$0.1; $^{**}$p$<$0.05; $^{***}$p$<$0.01. The base categories for dummy variables are:  ``Group 1'' for SUBDERE groups, ``False'' for Incumbent Mayor an ``Independent'' for Mayor. Only significant interactions are shown.} 
\end{tabular}
\end{table}

\begin{table}[!htbp] \centering 
\scriptsize
  \caption{OLS estimates for the full model (p-value RESET test = 0.3501) using all the municipalities, and the municipalities where CASEN is representative (139 municipalities). The third column shows the result of a bootstrap of 139 samples.} 
  \label{tab:SM_casen} 
\begin{tabular}{@{\extracolsep{5pt}}lD{.}{.}{-3} D{.}{.}{-3} D{.}{.}{-3} } 
\\[-1.8ex]\hline 
\hline \\[-1.8ex] 
 & \multicolumn{3}{c}{\textit{Dependent variable:}} \\ 
\cline{2-4} 
\\[-1.8ex] & \multicolumn{3}{c}{log (1 + ELAs)} \\ 
\\[-1.8ex] & \multicolumn{1}{c}{Full sample} & \multicolumn{1}{c}{CASEN sample} & \multicolumn{1}{c}{Bootstrap}\\ 
\hline \\[-1.8ex] 
 Log(population) & 0.778^{***} & 1.045^{***} & 0.815^{***} \\ 
  & (0.164) & (0.218) & (0.247) \\ 
 Higher education & 0.185^{***} & 0.284^{**} & 0.209^{**} \\ 
  & (0.065) & (0.124) & (0.102) \\ 
 Internet penetration rate & 0.103^{***} & 0.009 & 0.086 \\ 
  & (0.039) & (0.072) & (0.061) \\ 
    \textit{SUBDERE groups} &&&\\
  && \\[-1.8ex]
 \quad Group 2 & -0.122 & -1.254^{***} & -0.137 \\ 
  & (0.154) & (0.446) & (0.248) \\ 
 \quad Group 3 & -0.234 & -1.058^{***} & -0.263 \\ 
  & (0.196) & (0.370) & (0.300) \\ 
 \quad Group 4 & -0.129 & -1.215^{***} & -0.156 \\ 
  & (0.224) & (0.422) & (0.348) \\ 
 \quad Group 5 & 0.016 & -1.102^{**} & -0.050 \\ 
  & (0.277) & (0.472) & (0.422) \\ 
 \quad Group 6 & 0.297 & -1.153^{**} & 0.219 \\ 
  & (0.351) & (0.570) & (0.542) \\ 
 \quad Group 7 & 0.497 & -1.227^{*} & 0.405 \\ 
  & (0.417) & (0.628) & (0.652) \\ 
 \quad Group 8 & 0.236 & -1.458^{*} & 0.177 \\ 
  & (0.421) & (0.745) & (0.790) \\ 
 Log (community organizations) & 0.085^{*} & -0.010 & 0.075 \\ 
  & (0.051) & (0.093) & (0.089) \\ 
 Born in 1981 or after & -0.008 & 0.205^{*} & -0.025 \\ 
  & (0.056) & (0.119) & (0.085) \\ 
 Rurality & -0.052 & -0.076 & -0.042 \\ 
  & (0.054) & (0.139) & (0.090) \\ 
 Log (population density) & -0.139^{***} & -0.233^{**} & -0.114 \\ 
  & (0.053) & (0.102) & (0.105) \\ 
 Women & 0.066 & 0.236 & 0.012 \\ 
  & (0.069) & (0.269) & (0.144) \\ 
 Two-parent family (with children) & -0.093^{**} & -0.229^{***} & -0.094 \\ 
  & (0.043) & (0.072) & (0.068) \\ 
 Single-parent family (with children) & -0.101^{**} & -0.270^{**} & -0.057 \\ 
  & (0.043) & (0.107) & (0.100) \\ 
 Votes for current president & 0.148^{***} & 0.201^{***} & 0.125 \\ 
  & (0.046) & (0.075) & (0.130) \\ 
 Municipal officials & -0.040 & 0.111 & 0.015 \\ 
  & (0.170) & (0.398) & (0.275) \\ 
 Voter turnout & 0.087 & 0.196^{*} & 0.096 \\ 
  & (0.054) & (0.112) & (0.094) \\ 
 Mayor (government) & 0.063 & -0.137 & 0.020 \\ 
  & (0.069) & (0.155) & (0.125) \\ 
 Mayor (opposition) & -0.099 & -0.523^{**} & -0.084 \\ 
  & (0.079) & (0.248) & (0.169) \\ 
 Party affiliation & 0.161^{**} & 0.118 & 0.192 \\ 
  & (0.066) & (0.172) & (0.142) \\ 
 Incumbent mayor (True) & -0.020 & 0.140^{*} & -0.013 \\ 
  & (0.054) & (0.080) & (0.078) \\ 
 Evangelical Christians & -0.075^{***} & -0.017 & -0.071 \\ 
  & (0.027) & (0.064) & (0.048) \\
 Log (comm. org.) * Born in 1981 or after & -0.054 & -0.177^{*} &  -0.054\\  	
  & (0.043) & (0.100) & (0.075) \\ 
 Municipal officials * Voter turnout & 0.039 & 0.453^{**} & 0.067 \\  	
  & (0.095) & (0.224) & (0.206) \\ 
 Municipal officials * Mayor (opposition) & -0.027 & -1.123^{*} & -0.120 \\  		
  & (0.184) & (0.593) & (0.300) \\ 
 Constant & 0.037 & 1.092^{***} & 0.087 \\ 		
  & (0.209) & (0.418) & (0.327) \\ 
\hline \\[-1.8ex] 
Observations & \multicolumn{1}{c}{310} & \multicolumn{1}{c}{132} & \multicolumn{1}{c}{} \\ 
Adjusted R$^{2}$ & \multicolumn{1}{c}{0.814} & \multicolumn{1}{c}{0.853} & \multicolumn{1}{c}{} \\ 
Residual Std. Error & \multicolumn{1}{c}{0.424 (df = 275)} & \multicolumn{1}{c}{0.370 (df = 97)} & \multicolumn{1}{c}{} \\ 
F Statistic & \multicolumn{1}{c}{40.652$^{***}$ (df = 34; 275)} & \multicolumn{1}{c}{23.431$^{***}$ (df = 34; 97)} & \multicolumn{1}{c}{} \\ 
\hline 
\hline \\[-1.8ex] 
\multicolumn{4}{p{12cm}}{Note: $^{*}$p$<$0.1; $^{**}$p$<$0.05; $^{***}$p$<$0.01. The base categories for dummy variables are:  ``Group 1'' for SUBDERE groups, ``False'' for Incumbent Mayor an ``Independent'' for Mayor. Only significant interactions are shown.} 
\end{tabular} 
\end{table}

 \clearpage
 \newpage

\subsection{Concept selection and topic modeling}

\begin{table}
\scriptsize
  \caption{OLS regressions results for STM. Table shows the top three categories for each regression. Concepts in italic font were not included in the original list of concepts proposed by the government and were added by ELAs participants.} 
  \label{tab:coef_stm} 
\begin{center}
\begin{tabular}{l l l }
\hline \\[-1.8ex] 
 &  & \textit{Outcome variable: Topic} \\ 
\hline \\[-1.8ex] 
Topic 1: Environment &  Environmental respect / protection                   & $0.468 \; (0.008)^{***}$ \\
& \textit{Right to water}                                            & $0.295 \; (0.023)^{***}$ \\
& \textit{Conservation of cultural and historical heritage}       & $0.147 \; (0.055)$       \\
\hline \\[-1.8ex] 
Topic 2: Life & \textit{Respect life from conception}                      & $0.401 \; (0.047)^{***}$ \\
& Life                                              & $0.265 \; (0.004)^{***}$ \\
& Mental and physical integrity                        & $0.164 \; (0.007)^{***}$ \\
\hline \\[-1.8ex] 
Topic 3: Public policy &  \textit{Animal rights}                                        & $0.409 \; (0.044)^{***}$ \\
& Privacy and intimacy                                 & $0.100 \; (0.012)^{***}$ \\
& Right of association                                 & $0.099 \; (0.011)^{***}$ \\
\hline \\[-1.8ex] 
Topic 4: Unclassified & Election to public office                         & $0.102 \; (0.015)^{**}$  \\
& Decent housing                                    & $0.097 \; (0.004)^{***}$ \\
& Freedom to work                                      & $0.094 \; (0.013)^{**}$  \\
\hline \\[-1.8ex] 
Topic 5: Non - discrimination & Non - discrimination                              & $0.293 \; (0.008)^{***}$ \\
& Gender equity                                     & $0.105 \; (0.006)^{***}$ \\
& Equality                                          & $0.101 \; (0.004)^{***}$ \\
\hline \\[-1.8ex] 
Topic 6: Development & Free economic initiative / free enterprise        & $0.209 \; (0.013)^{***}$ \\
& \textit{Human Rights}                                      & $0.171 \; (0.014)^{***}$ \\
& Property                                          & $0.142 \; (0.009)^{***}$ \\
\hline \\[-1.8ex] 
Topic 7: Participation & Participation                                        & $0.457 \; (0.012)^{***}$ \\
& Suffrage / vote                                      & $0.424 \; (0.011)^{***}$ \\
& Election to public office                         & $0.407 \; (0.023)^{***}$ \\
\hline \\[-1.8ex] 
Topic 8: Rights & Judicial protection of individual rights          & $0.204 \; (0.008)^{***}$ \\
& \textit{Respect life from conception}                     & $0.189 \; (0.032)^{***}$ \\
& Life                                             & $0.145 \; (0.003)^{***}$ \\
\hline \\[-1.8ex] 
Topic 9: Security & Security / non-violence                              & $0.358 \; (0.005)^{***}$ \\
& Freedom of movement                                  & $0.112 \; (0.019)^{***}$ \\
& Decent housing                                    & $0.089 \; (0.004)^{***}$ \\
\hline \\[-1.8ex] 
Topic 10: Education & Education                                            & $0.336 \; (0.004)^{***}$ \\
& \textit{Right to quality public health care}               & $0.092 \; (0.014)^{***}$ \\
& Freedom of Education                              & $0.091 \; (0.007)^{***}$ \\
\hline \\[-1.8ex] 
Topic 11: Equality before the law & Equality before the law                           & $0.345 \; (0.005)^{***}$ \\
& Access to justice / due process                      & $0.286 \; (0.009)^{***}$ \\
& Equality                                        & $0.258 \; (0.005)^{***}$ \\
\hline \\[-1.8ex]
Topic 12: Social security & Social security                                  & $0.261 \; (0.005)^{***}$ \\
& Decent housing                                   & $0.257 \; (0.008)^{***}$ \\
& \textit{Right to work and a decent wage}                  & $0.148 \; (0.014)$       \\
\hline \\[-1.8ex]
Topic 13: Unclassified & Tax equality                                         & $0.437 \; (0.027)^{***}$ \\
& Equality in relation to public burdens               & $0.194 \; (0.042)^{***}$ \\
& Request before the authorities                       & $0.112 \; (0.022)^{***}$ \\
\hline \\[-1.8ex]
Topic 14: Indigenous people & Indigenous people                                    & $0.520 \; (0.011)^{***}$ \\
& \textit{Cultural identity of indigenous people}               & $0.506 \; (0.060)^{***}$ \\
& Cultural identity                                    & $0.471 \; (0.017)^{***}$ \\
\hline \\[-1.8ex]
Topic 15: Labor rights & Right to organize and to collective bargaining       & $0.413 \; (0.009)^{***}$ \\
& Right to strike                                      & $0.378 \; (0.013)^{***}$ \\
& \textit{Right to work and a decent wage}                  & $0.302 \; (0.016)^{***}$ \\
\hline \\[-1.8ex]
Topic 16: Freedom of education & Freedom of Education                              & $0.437 \; (0.012)^{***}$ \\
& Free economic initiative / free enterprise        & $0.304 \; (0.022)^{***}$ \\
& Property                                         & $0.286 \; (0.012)^{***}$ \\
\hline \\[-1.8ex]
Topic 17: Integration & Integration of disabled people                       & $0.243 \; (0.009)^{***}$ \\
& Equality before the law                         & $0.150 \; (0.005)^{***}$ \\
& Non - discrimination                             & $0.110 \; (0.004)^{***}$ \\
\hline \\[-1.8ex]
Topic 18: Childhood & Children and teenager's rights                       & $0.295 \; (0.007)^{***}$ \\
& \textit{Human Rights}                                  & $0.141 \; (0.012)^{***}$ \\
& Judicial protection of individual rights         & $0.132 \; (0.009)^{***}$ \\
\hline \\[-1.8ex]
Topic 19: Social rights & \textit{Social rights}                                        & $0.184 \; (0.009)^{***}$ \\
& Social security                                 & $0.129 \; (0.005)^{*}$   \\
& \textit{Standard of living}                                   & $0.117 \; (0.008)$       \\
\hline \\[-1.8ex]
Topic 20: Healthcare & Healthcare                                  & $0.296 \; (0.006)^{***}$ \\
& \textit{Right to quality public health care}              & $0.273 \; (0.018)^{***}$ \\
& Access to public information                         & $0.194 \; (0.011)^{***}$ \\
\hline \\[-1.8ex]
Topic 21: Freedom & \textit{Freedom of worship}                                 & $0.551 \; (0.017)^{***}$ \\
& \textit{Freedom of information and speech}                    & $0.435 \; (0.039)^{***}$ \\
& Freedom of expression                                & $0.411 \; (0.006)^{***}$ \\
\hline \\[-1.8ex]
Topic 22: Fair wage & Fair wage                                            & $0.261 \; (0.007)^{***}$ \\
& Gender equity                                    & $0.155 \; (0.007)^{***}$ \\
& \textit{Freedom of worship}                               & $0.128 \; (0.016)^{***}$ \\
\hline \\[-1.8ex]
Topic 23: Unclassified & \textit{Conservation of cultural and historical heritage} & $0.090 \; (0.039)$       \\
& Healthcare                                       & $0.081 \; (0.003)^{***}$ \\
& \textit{Right to quality public health care}              & $0.080 \; (0.009)^{***}$ \\
\hline \\[-1.8ex] 
\multicolumn{3}{l}{\scriptsize{$^{***}p<0.001$, $^{**}p<0.01$, $^*p<0.05$}}
\end{tabular}
\end{center}
\end{table}

\begin{figure}[!h]
\centering
\includegraphics[width=\textwidth]{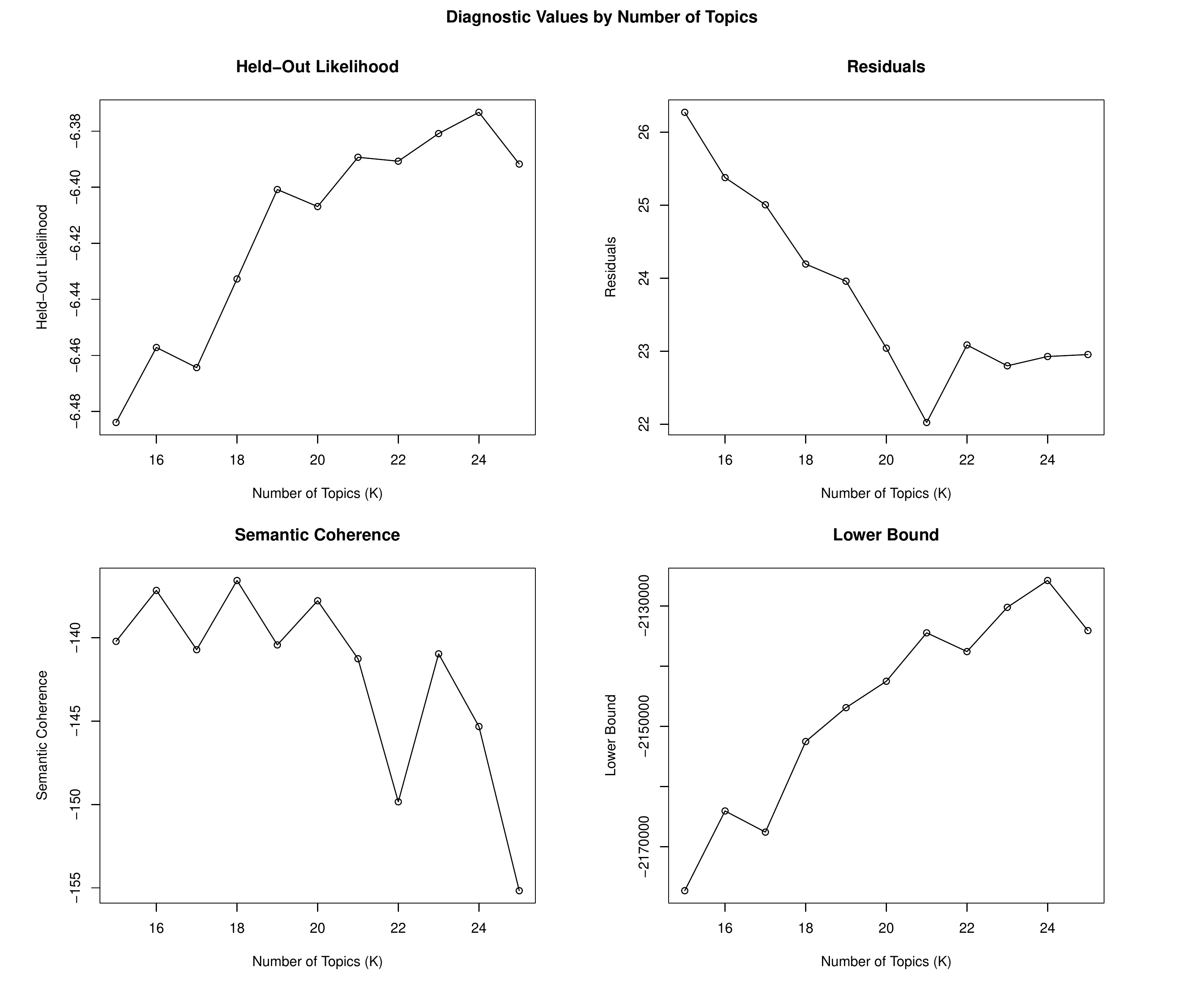}
\caption{Diagnostic values by number of topics. The optimal number of topics should seek to maximize the Held-Out Likelihood (top-left panel) and the Semantic Coherence (bottom-left panel), and minimize the residual dispersion (top-right panel). For 23 topics, both the Held-Out Likelihood and the Semantic Coherence are reasonably close to their maximum values, while  the residual dispersion has reached a stationary value (see \cite{wallach2009,taddy2012,mimno2011} for further information on these indicators).}
\label{fig:ntopic}
\end{figure}

\begin{figure}[!h]
\centering
\includegraphics[width=0.99\textwidth]{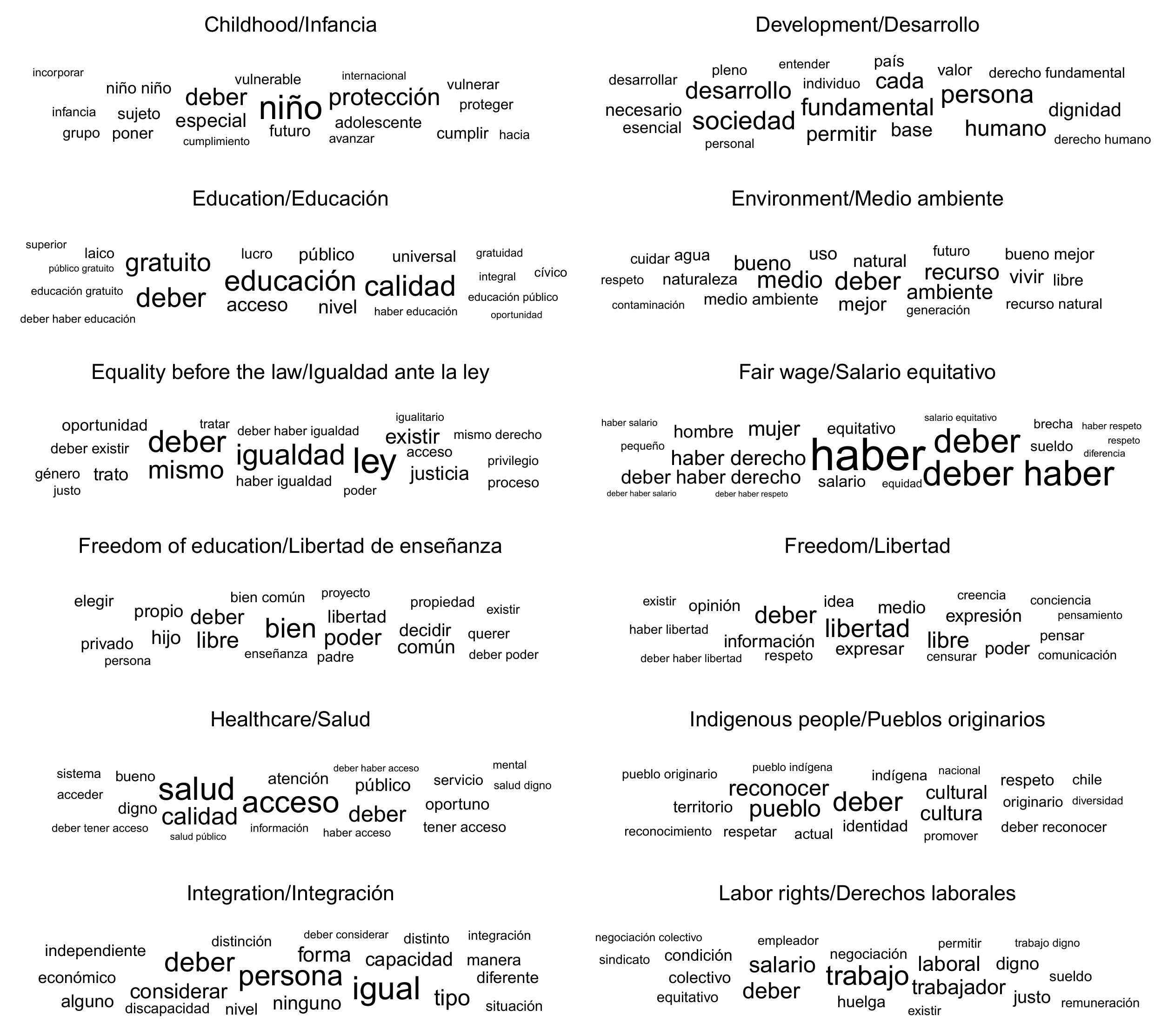}
\caption{Word Cloud display of highest probability words, for all topics.}
\label{fig:wordcloud1}
\end{figure}

\begin{figure}[!h]
\centering
\includegraphics[width=0.99\textwidth]{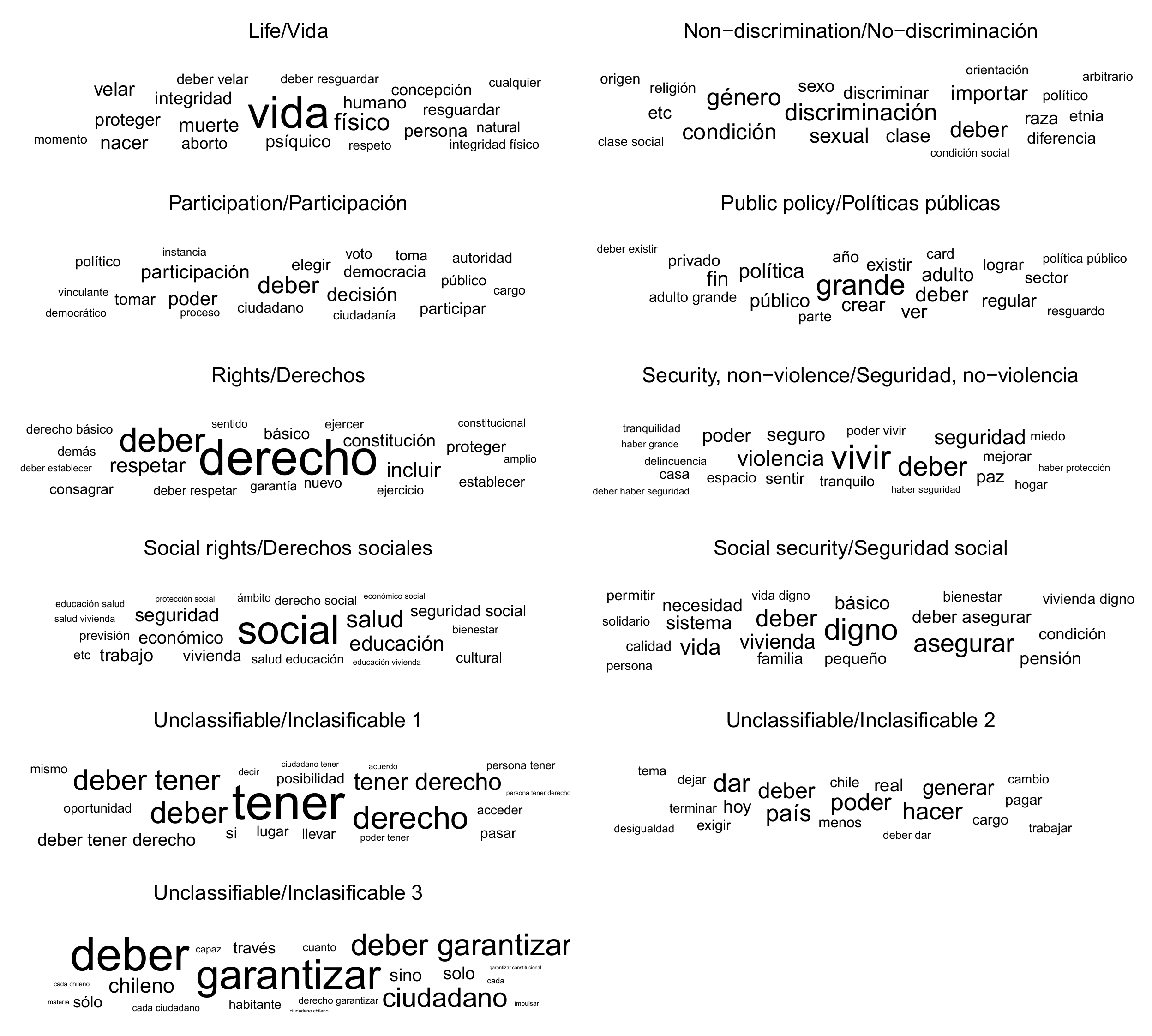}
\caption*{Word Cloud display of highest probability words, for all topics (continuation of Fig.~\ref{fig:wordcloud1}).}
\end{figure}

\begin{figure}[!h]
\centering
\includegraphics[width=0.99\textwidth]{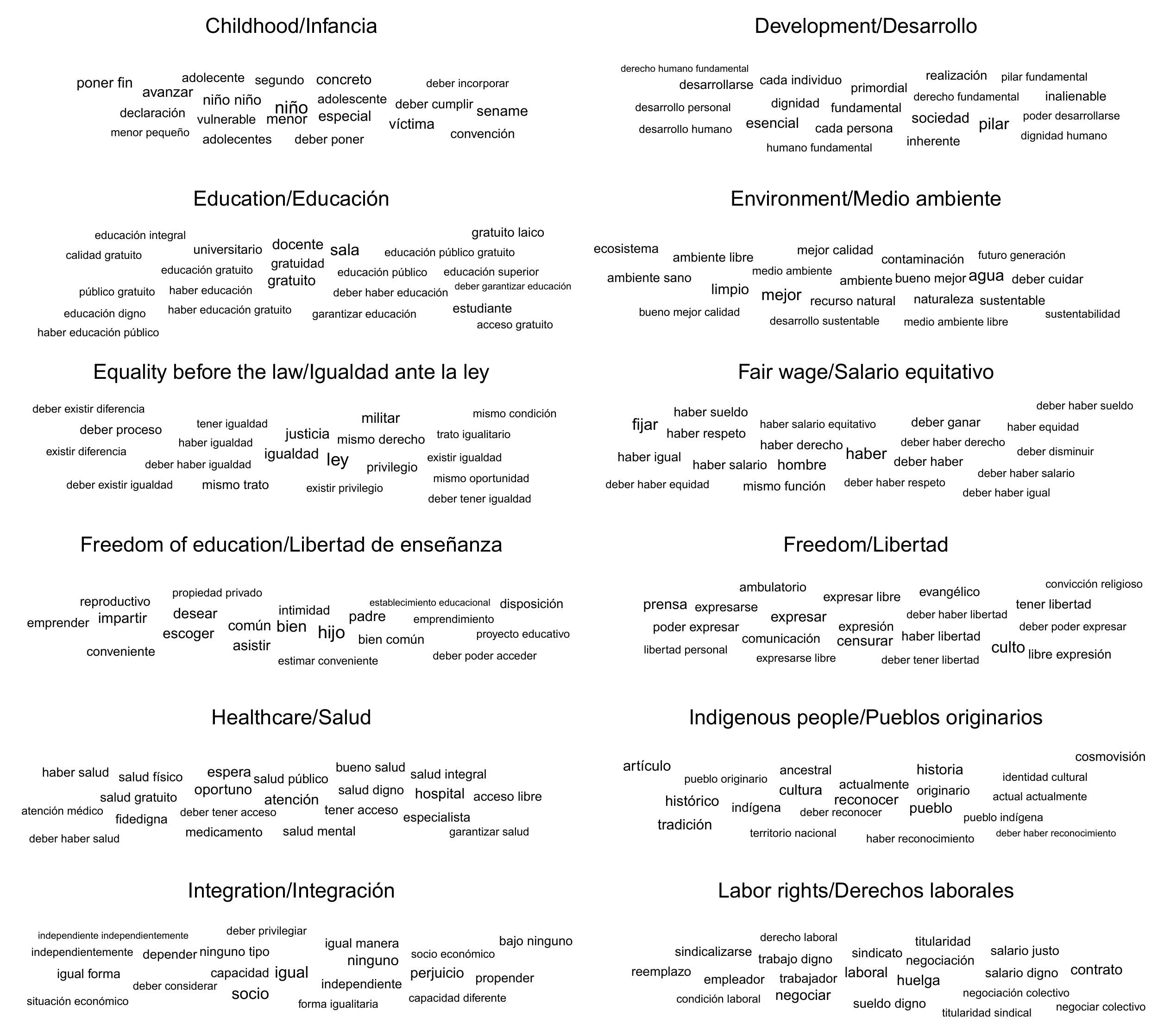}
\caption{Word Cloud display of frequent/exclusive words, for all topics.}
\label{fig:wordcloud055}
\end{figure}

\begin{figure}[!h]
\centering
\includegraphics[width=0.99\textwidth]{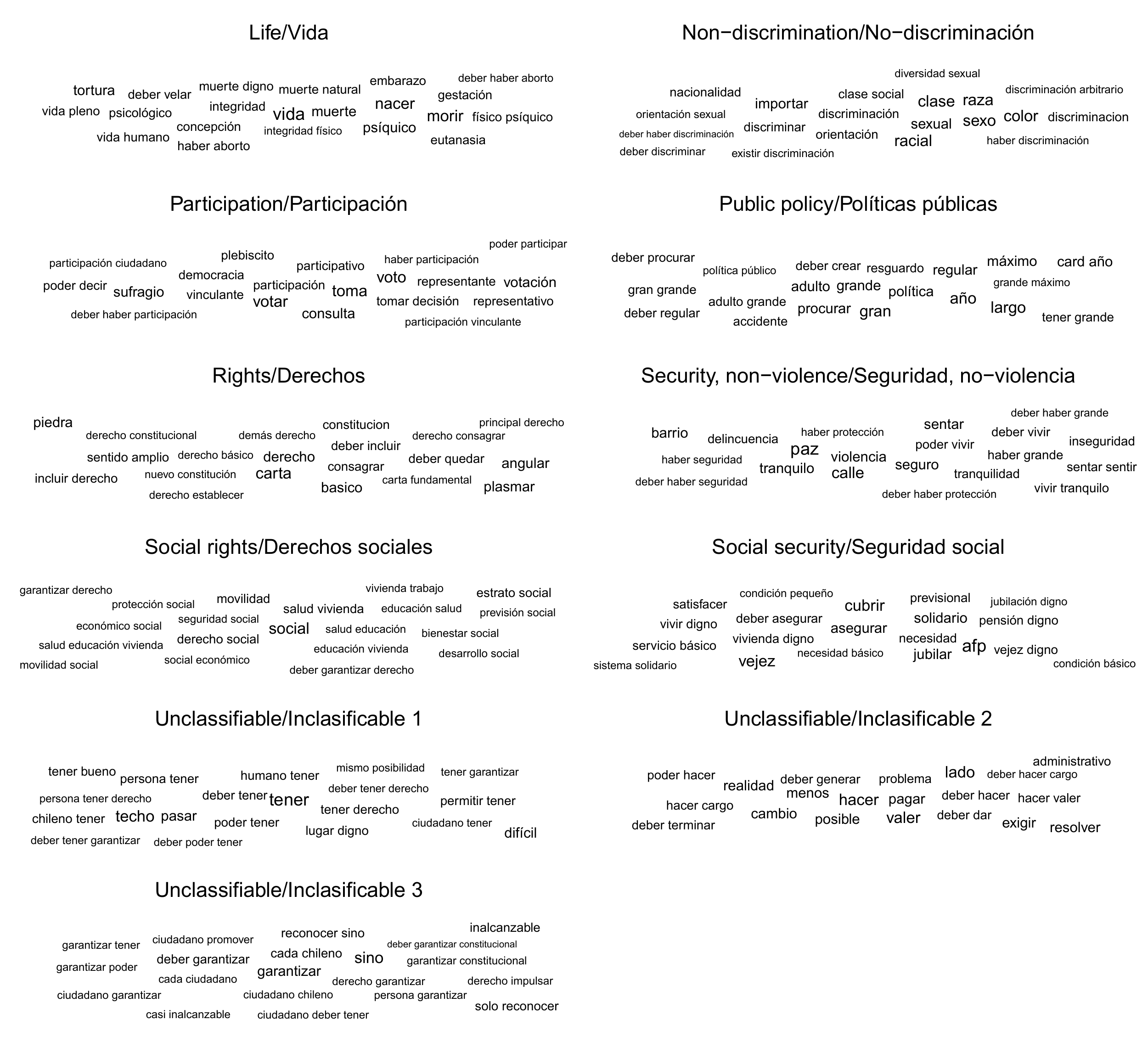}
\caption*{Word Cloud display of frequent/exclusive words, for all topics (continuation of Fig.~\ref{fig:wordcloud055}).}
\end{figure}

\pagebreak
\clearpage

\begin{table}[!htbp] \centering 
\scriptsize
  \caption{Topic: Environment} 
  \label{tab:stm_env} 
\begin{tabular}{@{\extracolsep{5pt}} l l L{3cm}  l L{6cm}} 
\\[-1.8ex]\hline 
\hline \\[-1.8ex] 
\multicolumn{1}{l}{Determinant} & \multicolumn{1}{l}{Quartile} & \multicolumn{1}{l}{Topic Words} & \multicolumn{1}{l}{City} & \multicolumn{1}{l}{Text} \\
\hline \\[-1.8ex] 
Votes & Bottom 25 & \multirow{5}{3cm}{recurso, medio ambiente, futuro, contaminación, sano, país, natural, libre, generación, recurso natural, uso, naturaleza, respeto.} &  Concón  & Debe poder incentivar la creación de empresas  entregando empleo y generando desarrollo. \\ 
 &  &  & Vitacura & Debe haber derecho a la vida en un medio ambiente libre de contaminación. \\ 
 &  &  & Lo Barnechea & Es fundamental para lograr tener una buena calidad de vida lo cual se logra a través del respeto de a la naturaleza y medio ambiente siendo labor de todos cuidarlo \\ 
 & & & & \\
\cline{2-5} \\[-1.8ex] 
 & Top 25 & \multirow{3}{3cm}{construir, deber, empresa, cuidar, mejor, bueno, agua.} &  Canela  & Tenemos derecho a disfrutar y tener acceso a los recursos naturales y cuidarlo para las futuras generaciones. \\ 
 &  &  & Puyehue & Debe asegurar recursos como son el agua aire tierra suelo y sub suelo. \\ 
 &  &  & La Pintana & Debemos mantenernos en equilibrio con la naturaleza para que las nuevas generaciones tenga un hábitat mejor pues se ha demostrado que gran parte de las enfermdades y catástrofes son responsabilidad de nosotros y las grandes empresas. \\ 
\hline \\[-1.8ex] 
Higher education & Bottom 25 & \multirow{3}{3cm}{agua, deber, mejor, bueno, actividad, tierra, naturaleza.}  &  Combarbabá & Debe nacionalizar y distribuir el agua   equitativamente regresando al estado y no siguiendo en manos de  los privados que lucran sin medida con ella ya que sin agua no hay vida. \\  
& & & Placilla & Debe dejar un mejor país a los hijos y nietos educando a los habitantes para que protejan la naturaleza. \\
& & & Collipulli & Debe tener un plan de contignecia ante las sequias que son comunes en nuestra región y localidad en específico habiendo derecho al agua para su consumo producción y sus animales. \\
\cline{2-5} \\[-1.8ex]  
& Top 25 & \multirow{5}{3cm}{cuidar, libre, cuidado, contaminación, generación, recurso natural, natural, futuro, recurso, sano, uso, medio ambiente, país.} &   Valdivia & Debe asegurar el respeto y cuidado por el medio ambiente en este derecho descansa el cuidado de nuestros recursos naturales.\\
& & & Valparaíso & Debe cuidar medio ambiente ya que la naturaleza forma parte de nuestras vidas responsables \\ 
& & & Antofagasta & Debemos conservar nuestro planeta sano.\\
\hline \\[-1.8ex]  
Pop. density & Bottom 25 & \multirow{2}{3cm}{agua, deber, respeto, país, cuidado, uso.} &  Chile Chico & Debe garantizar el respeto y protección del medio ambiente y la naturaleza.\\
& & & Tierra Amarilla & Debe cuidar aguas dulces sin contaminarlas. \\
& & & Pozo Almonte &  Debe haber respeto a todo lo que hace posible la vida en el planeta y a las condiciones para que pueda seguir existiendo la vida humana. \\
\cline{2-5} \\[-1.8ex] 
& Top 25 & \multirow{5}{3cm}{naturaleza, cuidar, contaminación, libre, fomentar, recurso natural, recurso, generación, natural, futuro, medio ambiente, sano, mejor, bueno.} &   Viña del Mar & Debe haber cuidado de la naturaleza preservando lo que existe y legislando para mejorar la protección del medio ambiente. \\
& & & Puente Alto & Debemos mantener el medio ambiente para las generaciones futuras.\\
& & & Talcahuano & Debe tener derecho a vivir en ambiente sano y un medio ambiente limpio. \\
\hline \\[-1.8ex]  
SEDI & Bottom 25 & \multirow{3}{3cm}{ambiente, naturaleza, agua, cuidar, mejor, natural, deber ,tierra.} &  Monte Patria & Debe tener ambiente mejor y más limpio para el futuro. \\
& & & Chanco & Debe cuidar de nuestros bienes naturales en la medida que no son renovables y cautelar la sobreexplotación.\\
& & & San Nicolás & Debe cuidar el medio ambiente pero de verdad con leyes que protejan la naturaleza sin distinción.\\
\cline{2-5} \\[-1.8ex]  
&  Top 25 &  \multirow{5}{3cm}{bueno, actividad, uso, recurso natural, habitante, fomentar, recurso, lograr, comunidad, manera, desarrollar, país.} & Calama & Deben ser usados administrados y explotados para el provecho de las comunidades en que se encuentran y del país como nación. \\
& & & Las Condes & Debe haber derecho a un medio ambiente sano y al desarrollo sustentable en donde se protejan los recursos naturales con una mejor regulación que evite su excesiva privatización.\\
& & & Providencia & Debe enfocar en la preservación de los recursos naturales y su impacto sobre las economías locales. \\
\hline \\[-1.8ex]  
\end{tabular}  
\end{table}

\begin{table}[!htbp] \centering 
\scriptsize
  \caption{Topic: Education} 
  \label{tab:stm_ed} 
\begin{tabular}{@{\extracolsep{5pt}} l l L{3cm}  l L{6cm}} 
\\[-1.8ex]\hline 
\hline \\[-1.8ex] 
\multicolumn{1}{l}{Determinant} & \multicolumn{1}{l}{Quartile} & \multicolumn{1}{l}{Topic Words} & \multicolumn{1}{l}{City} & \multicolumn{1}{l}{Text} \\
\hline \\[-1.8ex] 
Votes & Bottom 25 & \multirow{5}{3cm}{universal, acceso, público, oportunidad, laico, lucro, gratuito, integral, herramiento, educación pública, educación gratuita, calidad, educación.} &  Concón  & Debe garantizar una educación gratuita y de calidad para el desarrollo integral de las personas sin exclusión.  \\ 
 &  &  & Vitacura & Debe ser obligatoria gratuita y de calidad distinguiendo lucro de abuso.  \\ 
 &  &  & Lo Barnechea & Debe ser gratuita de excelencia e inclusiva un derecho social desde la primera infancia para lograr una igualdad de oportunidades en el futuro. \\ 
\cline{2-5} \\[-1.8ex] 
 & Top 25 & \multirow{4}{3cm}{superior, nivel, deber, educación cívica, deber haber educación, educación, gratis.} &  Canela  & Debe haber educación pública gratuita y de calidad para todos y en todos los niveles de educación. \\ 
 &  &  & Puyehue & Debe ser gratis y de buena calidad para todos/as especialmente los primeros 10 años. \\ 
 &  &  & La Pintana & Debe haber acceso a la educación en todos sus niveles: básica media institutos profesionales universitaria laica democrática multicultural no quedando ningún chileno sin educación por falta de recursos.  \\
\hline \\[-1.8ex] 
Higher education & Bottom 25 & \multirow{5}{3cm}{deber, docente, educación, educación cívica, nivel, cívica, conocimiento, deber haber educación.}  &  Combarbabá & Debe ser de libre acceso y de calidad en todos sus niveles y para todos. \\  
& & & Placilla & Debe encargar  el Estado  que los ciudadanos estén informados respecto de como funciona la sociedad en que vivimos debiendo considerar la educación cívica durante toda la vida para que sean los ciudadanos participantes de lo que se realice en su país. \\
& & & Collipulli & Debe haber educación de calidad y gratuita para todas y todos.\\
\cline{2-5} \\[-1.8ex]  
& Top 25 & \multirow{5}{3cm}{educación gratuita, integral, lucro, superior, laico, educación pública, público, universal, educación, calidad, gratuito, acceso.} &   Valdivia & Debe entregar una educación sin lucro laica gratuita de calidad asegurando el acceso a la cultura. \\
& & & Valparaíso & Debe ser completa integral y mas profunda sin limitarse sólo a la educación formal. \\ 
& & & Antofagasta & Debe haber educación abierta e igualitaria de calidad con acceso garantizado  a todos los chilenos y chilenas. \\ 
\hline \\[-1.8ex]  
Pop. density & Bottom 25 & \multirow{5}{3cm}{deber haber educación, profesor, deber, profesional, gratuidad, educación cívica, cívico, educación gratuita.} &  Chile Chico & Debe garantizar el acceso a una educación gratuita y de calidad con carácter público.\\
& & & Tierra Amarilla & Debe haber educación gratuita y no salir endeudados.. \\
& & & Pozo Almonte &  Debe ser ser gratuita y de calidad con profesores con vocación y bien remunerados. \\
\cline{2-5} \\[-1.8ex] 
& Top 25 & \multirow{5}{3cm}{acceso, gratuito, laico, lucro, herramienta, educación pública, universal, integral, nivel, público, calidad, educación.} &   Viña del Mar & Debe garantizar acceso a la educación de calidad en todos sus niveles.\\
& & & Puente Alto & Debe haber acceso equitativo gratuito y de calidad a la educación fortaleciendo a la educación pública y sus profesores. \\
& & & Talcahuano & Debe ser gratuita universal permitiendo que cada persona se eduque hasta donde sus capacidades lo otorguen.\\
\hline \\[-1.8ex]   
SEDI & Bottom 25 & \multirow{5}{3cm}{deber, sistema, docente, deber haber educación, educación cívico, educación digno, cívico, educación gratuito.} &  Monte Patria & Debe corregir abuso de alumnos frente al docente a los profesores se les han quitado atribuciones dentro del aula. \\
& & & Chanco & Es una buena educación de calidad para que las nuevas generaciones hagan de la nación un lugar mejor.\\
& & & San Nicolás & Debe existir una educación digna y de calidad para todos sin importar la calse social o ingresos económicos de las personas.\\
\cline{2-5} \\[-1.8ex]  
&  Top 25 &  \multirow{5}{3cm}{educación público, nivel, herramienta, integral, lucro, laico, educación, calidad, público, gratuito, universal, acceso.} & Calama & Debe otorgar acceso a una educación pública gratuita laica y de calidad que alcance hasta la educación universitaria. \\
& & & Las Condes & Debe haber educación de calidad desde la enseñanza parvularia hasta la universidad o educación superior y que sea sin lucro para las entidades que administran las instituciones.\\
& & & Providencia & Debe ser gratuita y de calidad en todos sus niveles con el deber de ingreso a la educación formal. \\
\hline \\[-1.8ex]  
\end{tabular}  
\end{table}

\begin{table}[!htbp] \centering 
\scriptsize
  \caption{Topic: Security} 
  \label{tab:stm_sec} 
\begin{tabular}{@{\extracolsep{5pt}} l l L{3cm}  l L{6cm}} 
\\[-1.8ex]\hline 
\hline \\[-1.8ex] 
\multicolumn{1}{l}{Determinant} & \multicolumn{1}{l}{Quartile} & \multicolumn{1}{l}{Topic Words} & \multicolumn{1}{l}{City} & \multicolumn{1}{l}{Text} \\
\hline \\[-1.8ex] 
Votes & Bottom 25 & \multirow{5}{3cm}{familia, espacio, vivir, poder, comunidad, lugar, seguro, entorno, paz, familiar, violencia, tranquilo.} &  Concón  & Debe cumplir requsitos básicos ya que es lugar donde se desarrolla la vida familiar que es el núcleo de nuestra sociedad.  \\ 
 &  &  & Vitacura & Debe consagrar la dignidad de las personas y el respeto asegurando en ellas el derecho a convivir en paz y tranquilidad.  \\ 
 &  &  & Lo Barnechea & Es agradable para todos los ciudadanos vivir en un estado seguro que permita a todos vivir tranquilos. \\ 
\cline{2-5} \\[-1.8ex] 
 & Top 25 & \multirow{4}{3cm}{poder vivir, mujer, sentir, deber, hogar, casa, seguridad, mejorar.} &  Canela  & Debemos vivir en paz conociendo y respetando los derechos y deberes; educando mejorando la comunicación en casa y a todo nivel cambiando los formatos de tv no más violencias; sino programas culturales educativos. \\ 
 &  &  & Puyehue & Debe buscar seguridad de la familia como fin. \\ 
 &  &  & La Pintana & Debe existir más vigilancia en las calles unión de vecinos y luminaria evitar violencia contra las mujeres y publicidad sexista.   \\
\hline \\[-1.8ex] 
Higher education & Bottom 25 & \multirow{3}{3cm}{mujer, mejorar, calle, tranquilo, deber, deber vivir, casa, poder vivir.}  &  Combarbabá & Merecemos tener un lugar donde vivir y criar a nuestros hijos un hogar digno y seguro. \\  
& & & Placilla & Debe dar seguridad inclusión e integración a la sociedad respetando la dignidad del ser humano. \\
& & & Collipulli & Debe ser derecho para todos el vivir tranquilamente y que nadie pase a llevar nuestra integridad física o psíquica.  \\
\cline{2-5} \\[-1.8ex]  
& Top 25 & \multirow{5}{3cm}{sentir, seguro, lugar, violencia, paz, familiar, seguridad, comunidad, vivir, poder, familia, espacio.} &   Valdivia &  Es vivir una vida sin violencia y que el estado garantice la seguridad de nuestros derechos. \\
& & & Valparaíso & Debe haber seguridad ciudadana dado que es importante para poder vivir y desarrollarnos en paz. \\ 
& & & Antofagasta & Debe sentir seguridad en el lugar en que uno este. \\
\hline \\[-1.8ex]  
Pop. density & Bottom 25 & \multirow{4}{3cm}{violencia, deber vivir, seguridad, casa, vivir tranquilo, entorno, deber, miedo, tranquilo.} &  Chile Chico & Debe garantizar la seguridad entendiéndola como un derecho social.\\
& & & Tierra Amarilla & Debe penalizar  a quienes agredan a personas física o psícamente  pues nadie puede irrespetar la integridad física y psíquica de los ciudadanos. \\
& & & Pozo Almonte & Debe proteger y resguardar a la mujer ya que hoy esta en desigualdad en muchos ámbitos como la violencia física salario. \\
\cline{2-5} \\[-1.8ex] 
& Top 25 & \multirow{5}{3cm}{poder vivir, vivir, poder, seguro, paz, lugar, sentir, hogar, comunidad, familia, espacio.} &   Viña del Mar & Debe poder vivir tranquilos en nuestro barrio y casa sintiendonos protegidos por la justicia y derechos. \\
& & & Puente Alto & Debe poder disfrutar de cada espacio público de manera segura garantizando vivir en un país sin violencia. \\
& & & Talcahuano & Debe procurar a todos los habitantes un medio seguro y una vida pacifica que asegure el desarrollo optimo de la comunidad y la persona.\\
\hline \\[-1.8ex]   
SEDI & Bottom 25 & \multirow{3}{3cm}{tranquilo, casa, poder vivir, poder, mejorar, calle, deber vivir, deber.} &  Monte Patria & Debe haber seguridad porque todos merecemos respeto y ya no podemos vivir tranquilos. \\
& & & Chanco & Debe contar con políticas de seguridad para evitar maltratos y robos.\\
& & & San Nicolás & Debe resguardar el derecho de seguridad permitiendo que las personas se desarrollen completamente aportando instancias en la sociedad.\\
\cline{2-5} \\[-1.8ex]  
&  Top 25 &  \multirow{5}{3cm}{miedo, sentir, paz, vivir, seguro, comunidad, familiar, violencia, lugar, seguridad, familia, espacio.} & Calama & Necesitamos una vida sin violencia y tranquilidad disfrutando en todo ámbito con espacios de seguridad. \\
& & & Las Condes & Debe estar seguro en espacios públicos y privados.\\
& & & Providencia & Debe erradicar el miedo estando libre de violencia de autoridades con paz y tranquilidad. \\
\hline \\[-1.8ex]  
\end{tabular}  
\end{table} 

\begin{table}[!htbp] \centering 
\scriptsize
  \caption{Topic: Equality} 
  \label{tab:stm_eq} 
\begin{tabular}{@{\extracolsep{5pt}} l l L{3cm}  l L{6cm}} 
\\[-1.8ex]\hline 
\hline \\[-1.8ex] 
\multicolumn{1}{l}{Determinant} & \multicolumn{1}{l}{Quartile} & \multicolumn{1}{l}{Topic Words} & \multicolumn{1}{l}{City} & \multicolumn{1}{l}{Text} \\
\hline \\[-1.8ex] 
Votes & Bottom 25 & \multirow{5}{3cm}{igualdad, ley, acceso, género, tratar, privilegio, proceso, deber, condición, diferencia, igual, oportunidad, deber haber igualdad.} &  Concón  & Debe haber igualdad en el acceso las oportunidades meritocracia y no ``pitutos'' y ante la ley.  \\ 
 &  &  & Vitacura & Somos iguales ante la ley al acceso a la justicia y al debido proceso.   \\ 
 &  &  & Lo Barnechea &  Es ser todos iguales ante la ley nadie puede ser tratado de manera diferente. \\ 
\cline{2-5} \\[-1.8ex] 
 & Top 25 & \multirow{3}{3cm}{trato, justicia, clase social, clase, mujer, mismo, hombre.} &  Canela  & Debe tener la posibilidad de defensa y trato de la misma forma ante la ley sin importar si tenemos o no plata. \\ 
 &  &  & Puyehue & Hay brecha de sueldos entre hombres y mujeres además la competencia beneficia a profesores y técnicos y no a obreros habiendo nula posibilidad laboral de transexuales.  \\ 
 &  &  & La Pintana & Deben ser tratados igual sin importar la clase social.  \\
\hline \\[-1.8ex] 
Higher education & Bottom 25 & \multirow{3}{3cm}{mujer, hombre, igual, deber, clase, deber haber igualdad, género.}  &  Combarbabá & No debe ser un obstaculo para el mundo laboral ya que el sexo femenino tiene las mismas capacidades laboras que el masculino. \\  
& & & Placilla & Debe cumplir este derecho de igual manera ante la ley independiente de la clase socio-económica o cargo. \\
& & & Collipulli & Debe haber igualdad de género  en la comunicación en el trato el respeto el trabajo y sueldo pues existe mayor valoración del hombre. \\
\cline{2-5} \\[-1.8ex]  
& Top 25 & \multirow{5}{3cm}{importar, tratar, existir, deber existir, privilegio, oportunidad, proceso, justicia, trato, mismo, diferencia, ley, igualdad.} &   Valdivia & No deben existir diferencias ni privilegios pues somos todos iguales.  \\
& & & Valparaíso & Entiende como un todo incluyendo la igualdad ante la ley de acceso a la justicia y al debido proceso ante las cargas públicas y proporcionalidad ante los tributos. \\ 
& & & Antofagasta & Debemos ser juzgados de igual manera con justicia imparcial. \\
\hline \\[-1.8ex]  
Pop. density & Bottom 25 & \multirow{3}{3cm}{importar, género, político, poder, deber, mismo derecho.} &  Chile Chico & Debe ser un salario sin distinción de género.\\
& & & Tierra Amarilla & Debe haber igualdad para sentirnos todos con los mismos derechos y oportunidades.  \\
& & & Pozo Almonte & Debe impulsar fuentes de trabajo eliminando el fuero político.  \\
\cline{2-5} \\[-1.8ex] 
& Top 25 & \multirow{5}{3cm}{privilegio, proceso, justicia, judicial, oportunidad, tratar, deber existir, acceso, ley, trato, existir, mismo, igualdad.} &   Viña del Mar & Debe haber igualdad ante género ante la ley acceso a la justicia y un debido proceso a desempeñar un cargo público etc siendo fundamentalmente al trato con igualdad y sin discriminación en su más amplio sentido.\\
& & & Puente Alto & Debe haber un mismo trato ante la ley poseyendo todos la misma condición. \\
& & & Talcahuano & Deben tener acceso a un proceso judicial que corresponda.\\
\hline \\[-1.8ex]   
SEDI & Bottom 25 & \multirow{5}{3cm}{género, deber, discriminación, existir, deber existir, existir igualdad, mismo oportunidad, juzgar, mismo derecho.} &  Monte Patria & Debemos ser individuos con las mismas oportunidades. \\
& & & Chanco & Debe dar oportunidades por igual a todas las personas sin discriminacion ni privilegios.\\
& & & San Nicolás & Debe garantizar la igualdad de género siendo necesario terminar con los abusos ya que hoy podemos afirmar con gran determinación que no existen diferencias entre distintos géneros.\\
\cline{2-5} \\[-1.8ex]  
&  Top 25 &  \multirow{5}{3cm}{deber haber igualdad, proceso, mismo, trato, privilegio, tratar, justicia, acceso, oportunidad, ley, igualdad.} & Calama & Debe contemplar las mismas normas o leyes para todas las personas sin que existan privilegios ni titulos nobiliarios. \\
& & & Las Condes & Debe haber igualdad  ante la ley acceso a la justicia y el debido proceso igualdad frente a tributos y cargas públicas.\\
& & & Providencia &  Debe tratar como igualdad de madios contemplando el concepto de equidad por el que no debe darse a todos lo mismo necesariamente. \\
\hline \\[-1.8ex]  
\end{tabular}  
\end{table} 

\begin{table}[!htbp] \centering 
\scriptsize
  \caption{Sentiment analysis for two variables of topic ``Security''.} 
  \label{tab:stm_sent} 
\begin{tabular}{@{\extracolsep{5pt}} L{3cm} R{15mm}  L{15mm}} 
\\[-1.8ex]\hline 
\hline \\[-1.8ex] 
word        &  value     &  sentiment \\
\hline \\[-1.8ex] 
\multicolumn{1}{l}{Socio-economic development} && \\
\hline \\[-1.8ex] 
tranquilo    &  98.68 \%  & Positive \\
casa         &  0.00 \%   & Neutral \\
poder vivir  &  96.71 \%  & Positive \\
poder        &  0.00 \%   & Neutral \\
mejorar      &  98.24 \%  & Positive \\
calle        &  0.00 \%   & Neutral \\
deber vivir  &  91.35 \%  & Positive \\
deber        &  0.00 \%   & Neutral \\
miedo        &  -94.96 \% & Negative \\
sentir       &  96.94 \%  & Positive \\
paz          &  98.24 \%  & Positive \\
vivir        &  96.08 \%  & Positive \\
seguro       &  97.81 \%  & Positive \\
comunidad    &  0.00 \%   & Neutral \\
familiar     &  96.94 \%  & Positive \\
violencia    &  -98.14 \% & Negative \\
lugar        &  0.00 \%   & Neutral \\
seguridad    &  97.81 \%  & Positive \\
familia      &  91.91 \%  & Positive \\
espacio      &  0.00 \%   & Neutral \\
\\[-1.8ex]\hline 
\hline \\[-1.8ex] 
\multicolumn{1}{l}{Primary Economic Activity} && \\
\hline \\[-1.8ex] 
establecer          &  0.00 \%    & Neutral \\
seguridad social    &  98.26 \%   & Positive \\
deber establecer    &  0.00 \%    & Neutral \\
ciudadano           &  0.00 \%    & Neutral \\
hogar               &  0.00 \%    & Neutral \\
paz                 &  98.24 \%   & Positive \\
seguridad           &  97.81 \%   & Positive \\
sentir              &  96.94 \%   & Positive \\
seguro              &  97.81 \%   & Positive \\
poder vivir         &  96.71 \%   & Positive \\
violencia           &  -98.14 \%  & Negative \\
deber haber grande  &  90.43 \%   & Positive \\
deber vivir         &  91.35 \%   & Positive \\
puerta              &  0.00 \%    & Neutral \\
delincuencia        &  0.00 \%    & Neutral \\
poder               &  0.00 \%    & Neutral \\
lugar               &  0.00 \%    & Neutral \\
vivir               &  96.08 \%   & Positive \\
\\[-1.8ex]\hline 
\hline \\[-1.8ex] 
\end{tabular}  
\end{table}


\end{document}